\documentclass{jaa}
\usepackage{graphicx}	
\usepackage{amsmath}	
\usepackage{amssymb}	
\usepackage{multirow}
\usepackage{array}
\usepackage{supertabular,booktabs}
\usepackage{adjustbox}
\usepackage{longtable}
\usepackage{pdflscape}
\usepackage{rotating}
\usepackage{lscape}
\usepackage{caption}
\usepackage{xcolor}
\usepackage{url}
\usepackage{lipsum}
\usepackage{longtable}
\usepackage[round]{natbib}
\usepackage{hyperref}
\hypersetup{colorlinks,citecolor=blue,urlcolor=blue,bookmarks=false,hypertexnames=true}
\usepackage{macro}

\newcommand{\sqdeg}{deg$^2$}

\def\mbh   {M$_{\rm BH}$ } 
\def\simi   {$\sim$\,}  
\def\z     {$z$ } 
\def\qj    {Q$_{\rm Jet}$ }

\def \deg         {\text{$^{\circ}$}}

\def \arcsec      {\text{$^{\prime\prime}$}}

\def \mujybeam    {$\mathrm{\muup}$Jy\,beam$^{-1}$}

\begin{document}\sloppy

\title{Decoding the giant extragalactic radio sources}


\author{Pratik Dabhade\textsuperscript{1}, D.J.Saikia\textsuperscript{2,3} and Mousumi Mahato\textsuperscript{2,4}}
\affilOne{\textsuperscript{1}Observatoire de Paris, LERMA, Coll\`ege de France, CNRS, PSL \& Sorbonne University, 75014, Paris, France\\}
\affilTwo{\textsuperscript{2}Inter-University Centre for Astronomy and Astrophysics (IUCAA), Pune 411007, India\\}
\affilThree{\textsuperscript{3}Department of Physics, Tezpur University, Napaam, Tezpur 784028, India\\}
\affilFour{\textsuperscript{4}Tartu Observatory, University of Tartu, Observatooriumi~1, 61602 T\~oravere, Estonia\\}


\twocolumn[{
\maketitle

\corres{pratik.dabhade@obspm.fr, pratikdabhade13@gmail.com}

\msinfo{9$^{\rm th}$ May 2022}{3$^{\rm rd}$ August 2022}

\begin{abstract}
Giant radio sources (GRSs) defined to be $>$ 0.7 Mpc are the largest single objects in the Universe and can be associated with both galaxies (GRGs) and quasars (GRQs). They are important for understanding the evolution of radio galaxies and quasars whose sizes range from pc to Mpc scales and are also valuable probes of their environment. These radio-loud active galactic nuclei (RLAGN) interact with the interstellar medium of the host galaxy on small scales and the large-scale intracluster or intergalactic medium for the GRSs. With several new and sensitive surveys over the last few years, the number of known GRSs has increased many fold which has led a resurgence of interest in the field. This review article summarises our current understanding of these sources based on nearly five decades of research, and discusses 
the importance of the Square Kilometer Array (SKA) in addressing some of the outstanding questions.
\end{abstract}

}]

include \year{2022} for  year of publication in the header

\doinum{12.3456/s78910-011-012-3}
\artcitid{\#\#\#\#}
\volnum{000}
\year{2022}
\pgrange{1--}
\setcounter{page}{1}
\lp{1}

\section{Introduction}\label{sec:1_intro}
One of the most magnificent phenomena associated with supermassive black holes in active galactic nuclei (AGN) is the  bipolar relativistic jets which originate in the vicinity of these black holes. Radio galaxies (RGs) and quasars (RQs), often collectively referred to as radio-loud AGN (RLAGN), range in size from less than about a few parsec (pc) for the most compact sources to hundreds of kpc for the largest sources, which extend to $\sim$5 Mpc. The extended radio structures of RLAGN have been classified into two major types: (i) edge darkened Fanaroff-Riley class I or FRI, and (ii) edge brightened Fanaroff-Riley class II or FRII  sources \citep{FR74}. The FRI sources have traditionally been found to be of lower luminosity and jet kinetic powers compared with FRII sources, the dividing luminosity being $\approx$10$^{26}$ W Hz$^{-1}$ at 150 MHz in 
a flat $\Lambda$CDM cosmological model based on the Planck results (H$_0$ = 67.8 km s$^{-1}$ Mpc$^{-1}$, $\rm \Omega_m$ = 0.308 \citealt{Plank2016}), which we adopt for this paper.
The radio jets in FRI sources tend to expand more rapidly forming diffuse plumes of emission and are more dissipative as they traverse outwards, while those in FRII sources remain well collimated forming bright hotspots, usually close to the outer edges of the lobes of radio emission. The hotspots are identified with the regions where the jets interact with the external medium dissipating most of its energy, while backflows from the hotspots form the extended lobes of radio emission. 

\begin{figure*}[ht!]
    \centering
    \includegraphics[width=12cm]{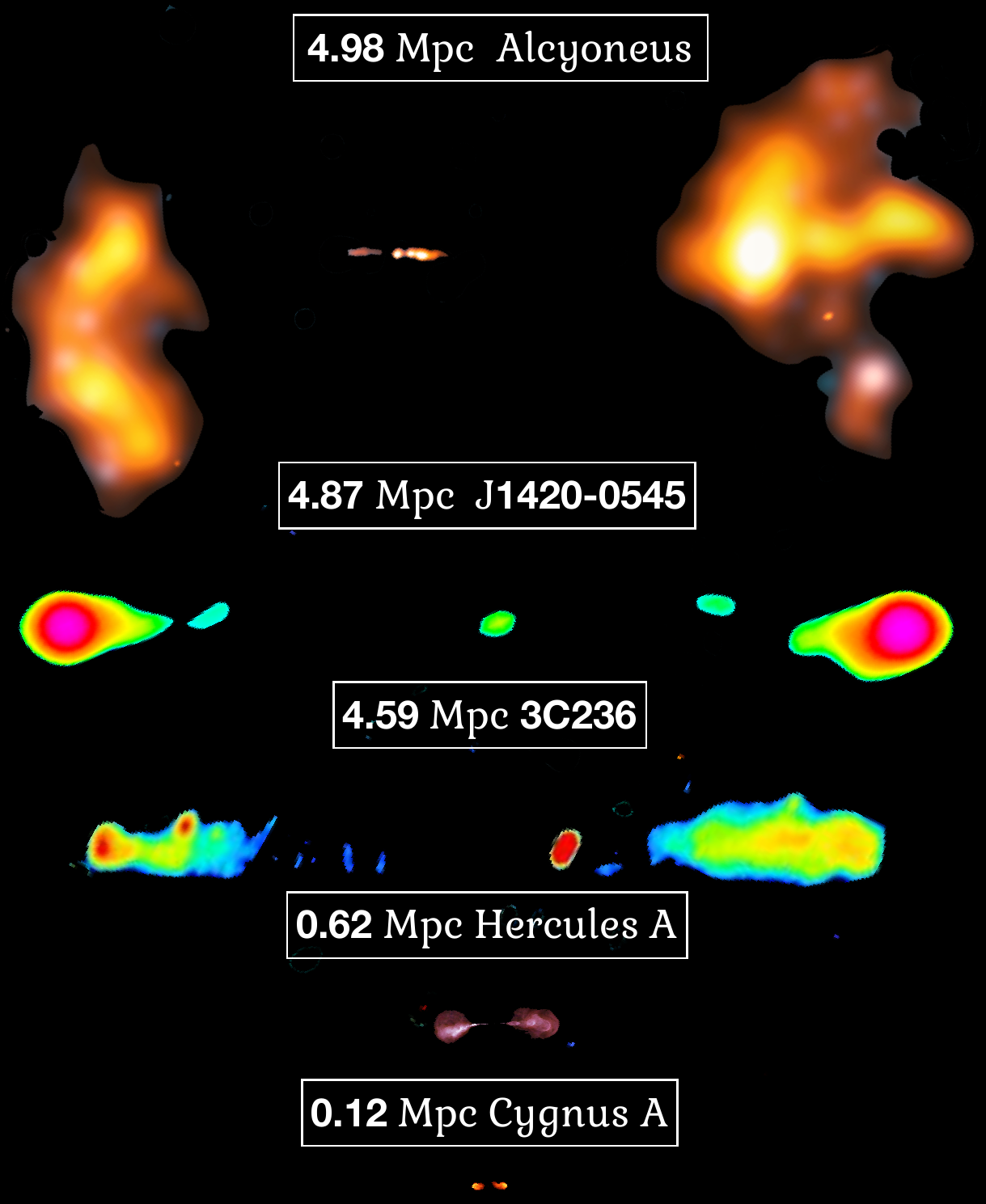}
    \caption{This figure illustrates how the biggest GRSs compare in projected linear sizes with Cygnus A and Hercules A, two of the well-known radio galaxies which were among the early ones to be identified with optical galaxies. All the images have been aligned horizontally for ease of comparison and this is not their true orientation. All the sources in the images have been scaled with respect to each other's sizes. The projected linear sizes have been computed using the cosmology mentioned in the Sec.~\ref{sec:1_intro} Credits:  \textit{Alcyoneus} with the image shown with two resolutions, 6\arcsec~ for the core and jets and 60\arcsec~ for the diffuse lobes - \citet{oeip122}; J1420$-$0545 - \citet{Machalski_2008}; 3C236 - \citet{mack97} and J.P Leahy's 3CRR atlas; Hercules A - (Credit: NASA, ESA, S. Baum and C. O'Dea (RIT), R. Perley and W. Cotton (NRAO/AUI/NSF)); Cygnus A - NRAO/AUI and \citet{Perley84}.}
    \label{fig:giantscale}
\end{figure*}

RGs have also been classified into low-excitation radio galaxies (LERGs) and high-excitation radio galaxies (HERGs) based on their optical spectra (e.g. \citealt{hardcastle07}; \citealt{bh12rgs}; \citealt{Heckman14}; \citealt{Tadhunter16}). The LERGs were found to exhibit an FRI structure although there are a number of FRII LERGs, while HERGs are predominantly FRIIs. The LERGs and HERGs 
appear to be linked to two different modes of accretion. In the low-excitation mode accretion is radiatively inefficient with an Eddington ratio ($\lambdaup_{\rm Edd}$) $<$1\%, while in the high-excitation mode $\lambdaup_{\rm Edd}$ is $>$1\% \citep{bh12rgs,Heckman14,Tadhunter16}. Recent studies by \citet{mingo19,Mingo22} have explored the relationships between FRII classes, HERGs, LERGs and properties of the host galaxies, and shown that these are related in subtle and interesting ways. 

\begin{figure*}[ht!]
    \centering
    \includegraphics[width=\textwidth]{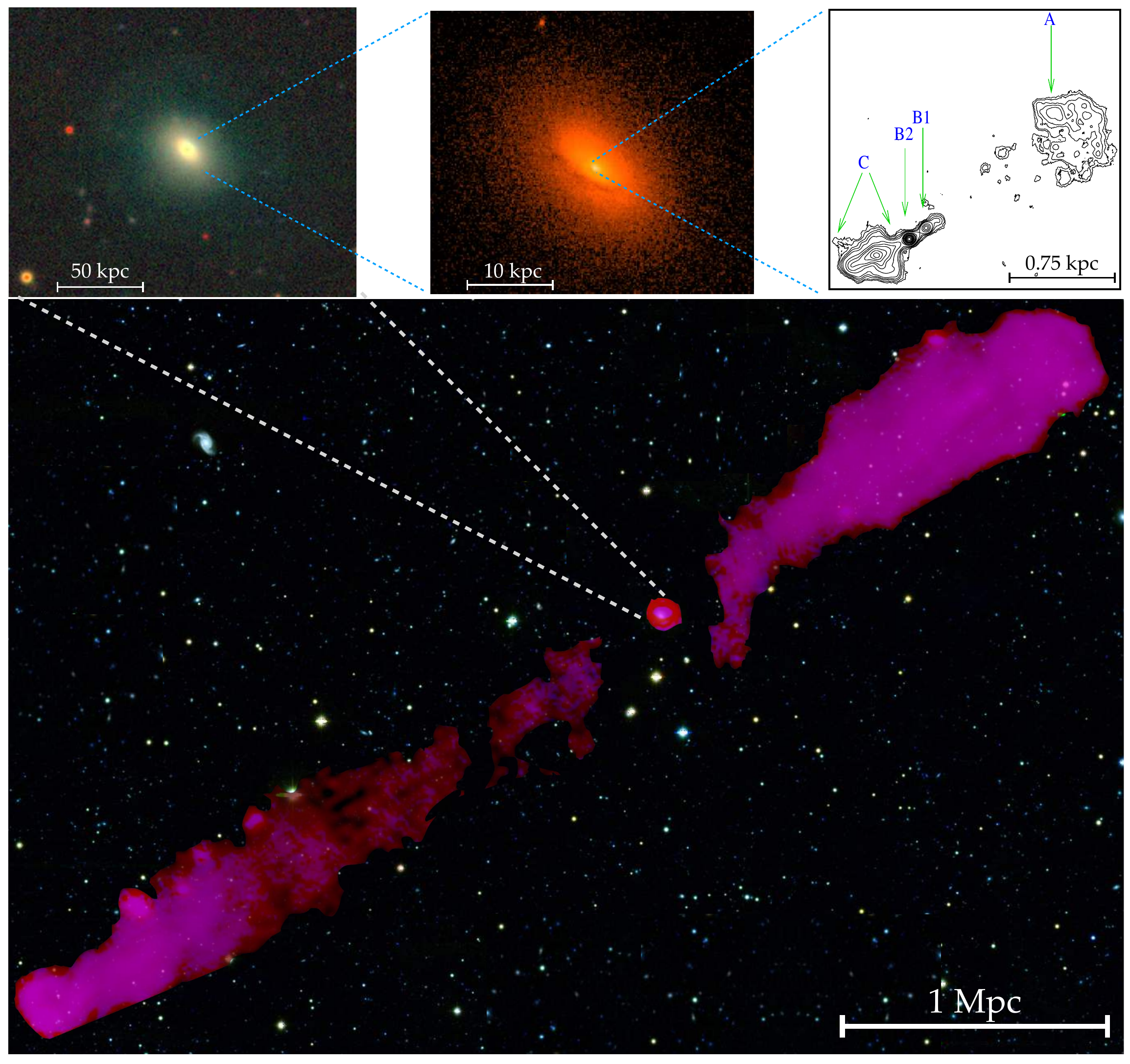}
    \caption{Multi-wavelength images of GRG 3C236 on various scales. Upper left: legacy survey DR9 optical composite image. Upper middle: high-resolution HST image of the optical galaxy. Upper right: VLBI mas scale image of the central radio component \citep{Schilizzi01}. Lower: composite image showing the LOFAR 144 MHz image \citep{Shulevski19} overlaid on an optical colour composite image from legacy survey DR9.}
    \label{fig:3c326ims}
\end{figure*}

In this review, we focus on giant radio sources (GRSs) which are defined to be $>$ 0.7 Mpc in size. The existence of such large sources was highlighted in the mid-1970s by \citet{willis74} who presented observations of 3C236 and DA240 with the Westerbork Synthesis Radio Telescope (WSRT; \citealt{Hogbom74}) with its high sensitivity and suitable resolution. GRSs may be associated with both radio galaxies and quasars; those associated with galaxies and quasars are hereinafter referred to as GRGs and GRQs, respectively. Similarly, we refer to smaller ($<$ 0.7 Mpc) sized radio sources, radio galaxies and radio quasars as SRSs, SRGs, and SRQs, respectively.
Both 3C236 and DA240 are associated with early-type galaxies. The GRSs are at the late stages of evolution of radio sources, and their large sizes are illustrated by comparing their projected linear sizes with two of the well-known radio galaxies which were among the early ones to be identified with optical galaxies, namely Cygnus A and Hercules A (Fig.~\ref{fig:giantscale}). Images of the GRG 3C236 \citep{willis74} which was the largest GRS till the discovery of J1420-0545 \citep{Machalski_2008} and more recently \textit{Alcyoneus} \citep{oeip122} are shown in Fig.~\ref{fig:3c326ims}. High-resolution VLBI observations of 3C236 have shown that the dominant central component which has a steep radio spectrum (spectral index\footnote{Defined as $S(\nu) \propto \nu^{-\alpha}$, where S$(\nu)$ is the flux density at a frequency $\nu$. We use this convention throughout this paper unless specified otherwise.} $\alpha >0.5$) is resolved into a double-lobed source with a one-sided jet \citep{Schilizzi01}. This was  one of the early examples of a rejuvenated radio galaxy showing signs of episodic activity, which is being discussed in Sec. \ref{sec:rejuv} From the early discovery of two GRGs by \citet{willis74}, the total number of GRSs known today is about 3200 and continues to grow. The progress in finding of new GRSs is discussed in Sec. \ref{sec:searchmain}.

\begin{figure}[ht!]
   \hspace{-0.75cm}
    \includegraphics[scale=0.4]{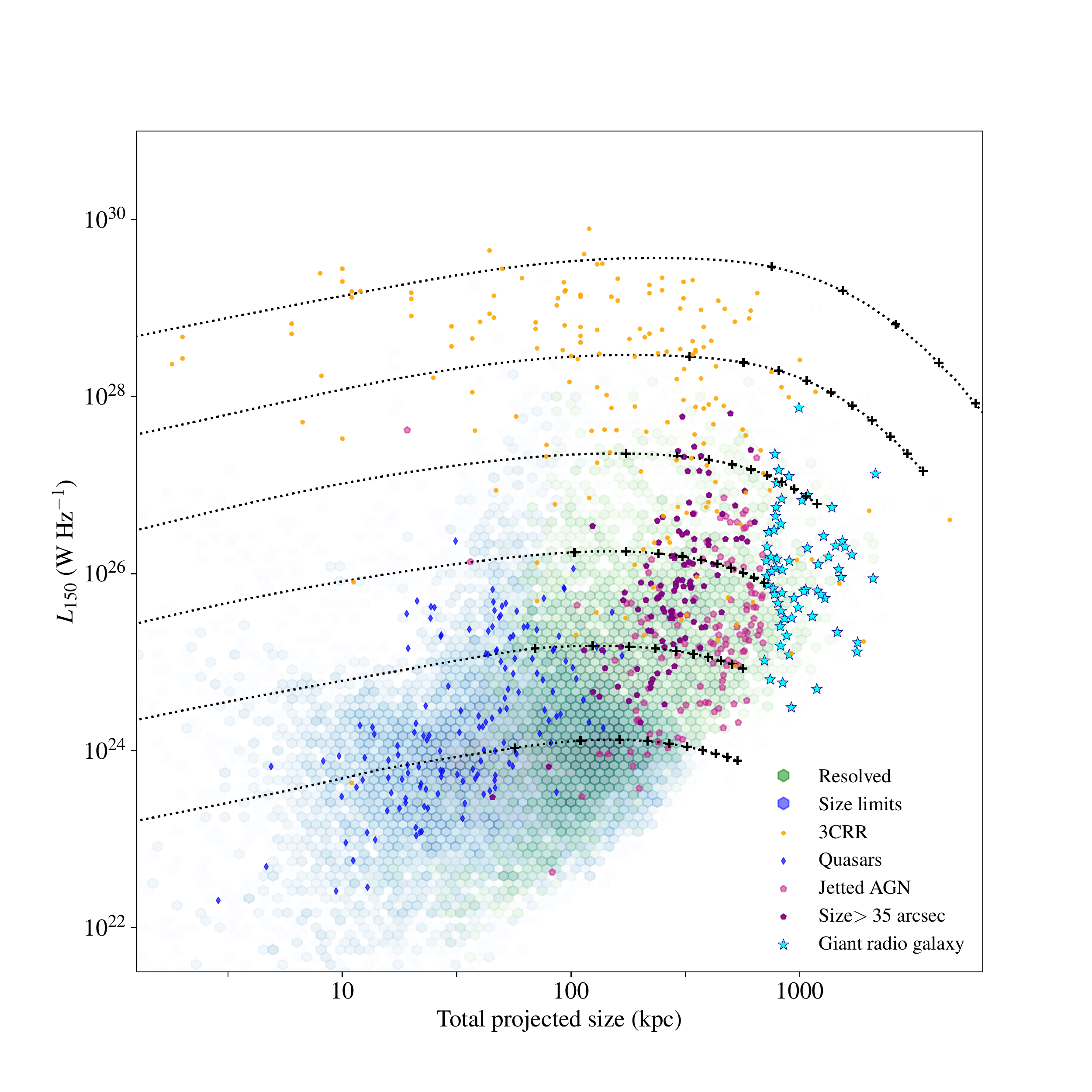}
    \caption{The radio power - projected linear size or P-D diagram for samples of sources from \citet{gurkan22}. The theoretical evolutionary tracks
    from \citet{hardcastle19} for $z$ = 0 sources lying in the plane of the sky in a group environment with mass within 500 kpc $= 2.5 \times 10^{13} M_\odot$ and kT $=$ 1 keV for two-sided jet powers Q = 10$^{35}$, 10$^{36}$,...,10$^{40}$ W increasing from bottom to top are shown in the Figure. Crosses on the tracks indicate time intervals of 50 Myr, so that each complete track is for 500 Myr. Image credit: \citet{gurkan22}.}
    \label{fig:pd}
\end{figure}

\subsection{Proposed models}\label{sec:models}
The GRSs represent the late stages of the evolution of radio galaxies and quasars and hence their formation and evolution are closely related to the evolution of smaller double-lobed radio sources. Over the years, 
several models have been explored to understand the evolution of radio sources as the jets propagate outwards in different environments from the nuclear regions to form these largest single objects in the Universe  \citep[e.g][]{rees71,longair73,Scheuer74,Blanford_Rees74,Begelman84,GK87,Kaiser97,Blundell99,Blundell00}. Our current understanding of RLAGN including their triggering, evolution and unification, accretion modes and feedback from AGN jets have been reviewed recently by \citet{Tadhunter16} and \citet{HardcastleCrostron20}.
The compact steep spectrum and peaked-spectrum sources, many of which evolve to the larger sources, have been reviewed by \citet{Odeasaikia21}. 

The radio power - projected linear size diagram or P-D diagram provides an important diagnostic to study the evolution of radio sources of different jet powers. The P-D diagram for different samples including the GRSs from the deep ASKAP (Australian SKA Pathfinder; \citealt{HOTANASKAP21}) EMU (Evolutionary Map of Universe; \citealt{EMUNORIS21}) Survey of the GAMA23 field from \citet{gurkan22} is shown in Fig.~\ref{fig:pd}. The figure also shows the evolutionary tracks of $z=0$ sources lying in the plane of the sky in a group
environment with two-sided jet power increasing upwards  10$^{35}$, 10$^{36}$,~...~,10$^{40}$ W \citep{hardcastle19,gurkan22}. There is considerable scatter in the diagram because of different jet powers, different environments in which the sources are evolving, and different angles of inclination of the source axes to the line of sight which will affect the degree of relativistic beaming and also the projected linear sizes. At low flux densities and low luminosities, the radio emission could be dominated by star formation activity rather than an AGN. Considering the RLAGN, the GRSs represent the end-stages of the evolution of RLAGN and most are in the range of jet power from 10$^{36}$ to 10$^{39}$ W. The crosses in Fig.~\ref{fig:pd} are in intervals of 50 Myr so that each complete track is for 500 Myr. As expected sources with high-power jets reach the Mpc-scales on a shorter time scale than low-power ones. 

However although the total number of GRSs is about \simi3200 (see Tab.\ \ref{tab:radiosur} and Sec. \ref{sec:searchmain}),
it constitutes only a small fraction of the total population of radio galaxies and quasars in well-observed samples. Therefore it is essential to enquire what conditions lead to the formation of GRSs. A number of possibilities have been explored as listed below.

\begin{enumerate}
\item The growth of GRSs is favoured in sparse or low-density environments \citep{mack98,malarecki15}.
\item The AGNs of GRSs are extremely powerful \citep{gk89} and hence these are able to
produce exceptionally powerful radio jets when compared with normal radio galaxies, leading to their large sizes \citep{wita-grg-agn}.

\item Based on the evidence of discontinuous beams in GRSs, \citet{ravi96} suggested that the GRSs could be associated with the recurrent activity of AGN or multiple epochs of activity.
\end{enumerate}
It is quite plausible that the formation of GRSs is not due to one single factor but a combination of factors such as powerful AGN with high jet power, fueling, accretion processes and the external environment in which the source evolves. We comment further on these aspects in the concluding section (Sec.\ \ref{sec:concl}) of the article.

\begin{table*}[ht!]
\centering

\setlength{\tabcolsep}{9.0pt}

\caption{The table presents the number of GRSs found from different major radio surveys (arranged according to increasing frequency (MHz) of observations) with approximate sky areas and sensitivities. Except for LoTSS GRSs, many of the GRSs required observations at other radio frequencies to confirm their classification as a GRS. For 3CRR$\dagger$ we use the online \href{https://3crr.extragalactic.info}{database}. To avoid counting a GRS multiple times, it has been counted against the first paper which reported it. Column (1): Survey name ; Column (2): Observing frequency of the survey ; Column (3): Sky area covered by the survey; Column (4): rms of the survey ;  Column (5): Number of new GRSs reported in the given reference/survey; Column (6): References. 
* They define GRSs as $>$ 1 Mpc.}\label{tab:radiosur}

\begin{tabular}{lccccr}
\hline
 Survey & Frequency & Sky-Area  & rms   & N  & References\\ 
        & (MHz)     & (\sqdeg)  & (mJy~beam$^{-1}$)    & &\\ 
   (1)     & (2)     & (3)  & (4)   & (5) & (6)\\ 

\hline
LoTSS-DR1 & 144 & 424      & 0.07         & 206 &\cite{PDLOTSS} \\
LoTSS-Bo{\"o}tes     & 144 & 26.5    & 0.03-0.07    & 74  & \citet{Simonte2022}\\
LoTSS-DR2 & 144 & 5700     & 0.08        &     2050 & \citet{oeidr2}\\
LOFAR-H-ATLAS & 150 &  142    & 0.1-2            & 6     &\citet{Hardcastle2016} \\
7C        & 151 & 473      &15          &  14    & \citet{cotter96} \\
3CRR$\dagger$      & 178 & 15287    &2000            &12    & \cite{3crr}            \\
WENSS     & 325 & 10176    & 3         &  31     & \cite{Schoenmakers2001} \\
MRC       & 408 & 1937     &70            & 8     & \cite{kapahi98}       \\
SUMSS     & 843 & 2100     &  1              & 17   &\cite{Saripalli2005}  \\
RACS$^{*}$      & 888 & 1059     &  0.25        & 178 &\cite{Andernach21} \\
EMU       & 888 & 83       & 0.04         & 63  & \cite{gurkan22} \\
EMU$^{*}$       & 1000 &30       & 0.03        & 19  & \citet{bruggen21}\\
NVSS      & 1400& 1543    &0.45             & 21   & \cite{Machalski_2001} \\
NVSS      & 1400& 4563    &0.45             & 34   & \cite{Machalski_2007} \\
NVSS      & 1400& 10308    &0.45            & 30   & \cite{lara2001} \\
NVSS      & 1400& 34000    &0.45            & 9    & \cite{Solovyov14} \\
NVSS      & 1400& -    &0.45            & 80   & \cite{Amirkhanyan2016} \\
NVSS      & 1400& -    &0.45           & 22   & \cite{D17} \\
NVSS      & 1400& -    &0.45            & 148  & \cite{sagan1} \\
NVSS      & 1400& 14555    &0.45           & 24   & \cite{koziel20} \\
NVSS      & 1400& 9376    &0.45            & 134   & \cite{kuzmic21} \\
NVSS      & 1400& -    &0.45             & 10   & \cite{sagan3} \\

\hline
\end{tabular}
\end{table*}

\begin{figure*}[ht!]
    \centering
    \includegraphics[width=\textwidth]{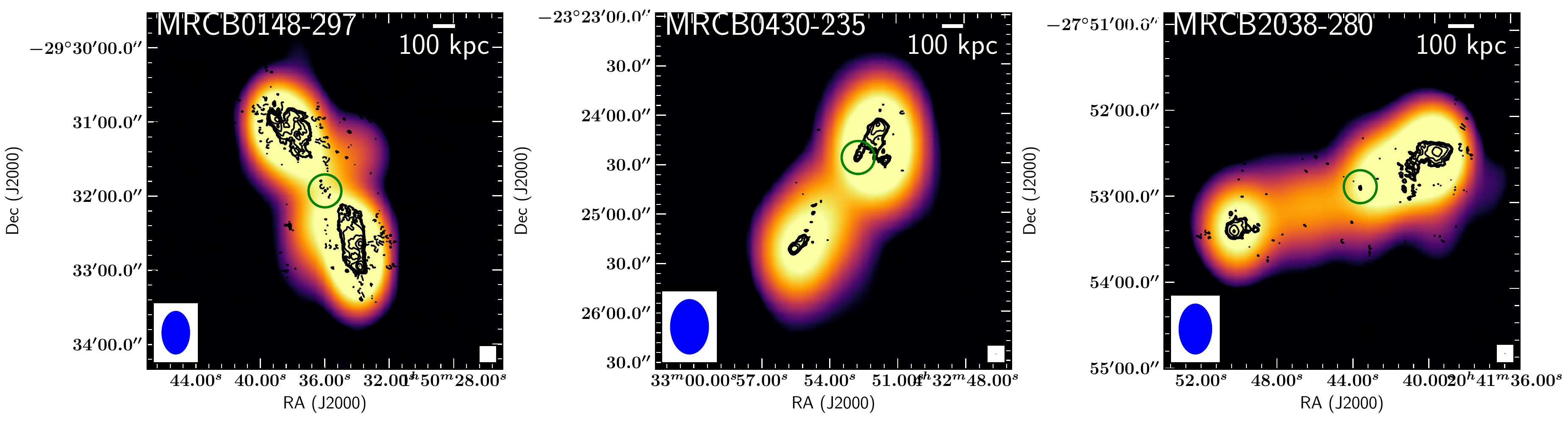}
    \caption{Three newly identified GRGs from the MRC as discussed in Sec.\ \ref{sec:newgrgs} and listed in Tab. \ref{tab:mrc} are shown here. The black contours from VLASS are overlaid on TGSS radio maps shown in colour. The location of the host galaxy is identified with a green marker. The beams of TGSS and VLASS are shown on the bottom left (blue) and right (black) corners, respectively. Further details are given in Tab.\ \ref{tab:mrc}.}
    \label{fig:mrcgrs}
\end{figure*}

\vspace{-0.6cm}
\section{GRS searches and present status} \label{sec:searchmain}
The discovery of new GRSs after \citet{willis74} highlighted the Mpc-scale structure of 3C236 and DA240,
progressed gradually in the initial years. There were only 53 GRSs in the compilation by \citet{Ishwara1999} 25 years later.
Until the advent of deeper radio surveys over large areas of the sky like the NRAO VLA Sky Survey (NVSS; \citealt{nvss}) and Westerbork Northern Sky Survey (WENSS; \citealt{wenss97}) and more recently the LOFAR Two Metre Sky Survey  (LoTSS; \citealt{lotssshimwell,lotssdr2}), where LOFAR stands for LOw Frequency ARray \citep{lofar}, and the ASKAP surveys \citep{racs20,EMUNORIS21,gurkan22}, GRSs were mostly found using targeted follow up radio observations of extended sources from strong-source catalogues (e.g. 3CR, B2, and PKS).
Targeted searches for GRSs from surveys like 7C, NVSS, and WENSS yielded many GRSs and promising GRS candidates. Owing to the relatively coarser resolutions of these surveys, follow up high-resolution radio observations were often needed to establish the GRS nature of these sources. Detection of the radio cores helped make reliable optical identifications. The finding of new GRSs
was also greatly aided by surveys of the sky at optical wavelengths (e.g. Sloan Digital Sky Survey, SDSS, \citealt{sdss00}; 
and 6 degree Field Galaxy Survey, 6dFGS, \citealt{6dFJones}) which along with the high-resolution radio observations  
provided reliable optical identifications and redshifts. 

These developments led to a surge in interest in finding new GRSs and studying the properties of these objects using large samples along with using newer multi-wavelength data from various 
surveys.
\cite{Kuzmicz2018} presented a catalogue of 349 GRSs based on their compilation up to 2018 along with the properties. Later, \cite{sagan1} presented a catalogue with double the number of GRSs reported in \cite{Kuzmicz2018}, owing to the addition of a large sample from \citet{PDLOTSS}. Together, these two compilations provide references to GRSs reported till March 2020. Since then, more GRSs have been reported and studied \citep{Tang20,ishwar20,Galvin20,Bassani21,Delhaize21,Andernach21,Masini21,kuzmic21,bruggen21,sagan3,oeip122,gurkan22,Simonte2022,oeidr2}. Further details on the discoveries of GRSs over the years can be found in the introduction of \citet{pdthesis}.

A few of the major radio surveys from which GRSs have been identified over the years are listed in Tab.\ \ref{tab:radiosur}. These include the 3CRR \citep{3cr,3crr}, Molonglo Reference Catalogue (MRC; \citealt{large82mrc,kapahi98}), WENSS \citep{wenss97,Schoenmakers2001}, Sydney University Molonglo Sky Survey (SUMSS; \citealt{sumss99,Saripalli2005}), Rapid ASKAP Continuum Survey (RACS; \citealt{racs20}), EMU survey \citep{EMUNORIS21}, NVSS by a number of authors (see Tab.\ \ref{tab:radiosur}) and recent LOFAR surveys \citep{lotssshimwell,lotssdr2} which have contributed most of the presently known GRSs. The total number of GRSs listed in Tab.\ \ref{tab:radiosur} is 3190. The number of GRSs have been listed after verifying that these belong to this category in the current cosmological model and may differ from the original papers listed. Also to avoid counting a GRS multiple times, it has been attributed to the authors who first reported it. In Tab.\ \ref{tab:radiosur} we list papers which have reported at least 6 new GRSs from the surveys. Including GRSs not listed in the Tab.\ \ref{tab:radiosur}, there are approximately 3200 known GRSs.

The LOFAR Two-metre Sky Survey \citep[LoTSS;][]{lotssshimwell} at 144~MHz, is one of the most sensitive radio surveys at low frequencies over a large area of the sky, reaching noise levels of \simi~70~\mujybeam with the best resolution of \simi6\arcsec. Using the LoTSS data release 1 (DR1), \citet{PDLOTSS} found 239 GRSs of which 225 were new from a sky area of 424~\sqdeg. Now, using LoTSS-DR2, 1038 new GRSs have been found from \simi5700~\sqdeg \citep{oeidr2}.
Therefore, it is clear that with better sensitivities, especially at low radio frequencies more GRSs will be found. The number of GRSs is likely to be higher than the reported ones as there are candidate GRSs whose size requires confirmation by finding their host galaxies and determining their redshifts. For example, as LoTSS-DR2 covers an area which is about 13 times larger than LoTSS-DR1 and has been observed with similar sensitivity and better dynamic range (see section 3.6 in \citealt{lotssdr2}), the number of GRSs in this field is likely to be larger than what has been found so far. Many of the GRSs in these sensitive surveys could be at high redshifts ($\gtrsim$ 1.5). Dedicated spectroscopic optical observational programs are needed to confirm their nature. 

However, although the number of known GRSs has increased significantly, they constitute only a small fraction of the sources in the surveys. For example, consider the sources in LoTSS-DR1 and LoTSS-DR2 from which we have the largest number of GRSs. LoTSS DR1 contains 325694 radio sources, has a source density of 770 sources per square degree and a point-source completeness of 90\% at an integrated flux density of 0.45 mJy. The LoTSS-DR2 catalogue consists of 4396228 radio components from their 6 arcsec resolution total intensity (Stokes I) maps where the median rms sensitivity is 83 $\muup$Jy~beam$^{-1}$ and the point-source completeness is 90\% at a peak brightness of 0.8 mJy~beam$^{-1}$. Although the weak radio sources consist largely of star-forming galaxies \citep{Richards99,mauchsadler07,Sabater19}, the fraction of known GRSs from these surveys is extremely small.

\subsection{Alternate new search methodology}
In the last decade, several large-scale radio surveys have been made using different radio telescopes, providing a huge amount of data. Also, in future, this flow of new deep radio data is going to increase with more surveys, especially from the SKA\footnote{\url{https://www.skao.int/en}}. Hence, it has become nearly impossible to manually analyze and inspect data and this has led to alternate or new efforts to keep up with the fast data flow.

Throughout astronomy, automated techniques such as artificial neural network (ANN),  machine learning (ML), and deep learning have shown their capabilities in classifying and finding sources. Similarly, a few works \citep{proctorGRS,Galvin20,Tang20,Tang22} have successfully demonstrated its usefulness in identifying and classifying radio sources or GRSs with limited human intervention. Clearly, the SKA-era astronomy will be dominated by its usage.

A possible alternative to the above technique is the citizen science approach, which employs hundreds of trained human users to manually inspect a large amount of data. In radio astronomy, there have been similar efforts via the Radio galaxy zoo\footnote{\url{https://radio.galaxyzoo.org}} project, the RAD@home\footnote{\url{https://www.radathomeindia.org}} project \citep{Hota16}, and the radio galaxy zoo:LOFAR\footnote{\url{https://www.zooniverse.org/projects/chrismrp/radio-galaxy-zoo-lofar}}, which has also led to the finding of a few GRSs \citep[e.g.][]{Banfield15,kapinska17}.

\subsection{New GRGs}\label{sec:newgrgs}
As part of the work for the current paper, we carried out an extensive literature review and as a result, were able to find new GRSs which were previously not reported as giants from the 1~Jy MRC sample \citep{kapahi98}. The three newly recognised sources are listed along with their basic properties in Tab.\ \ref{tab:mrc}. Their radio images are presented in Fig.\ \ref{fig:mrcgrs}, where TGSS  \citep[TIFR-GMRT Sky Survey;][]{tgss_intema} maps are overlaid with high resolution VLASS (VLA Sky Survey; \citealt{Lacy2020VLASS,vlass21}) contours. The images clearly show all three of them to be associated with galaxies and have an FRII type of structure.

\begin{table*}[htb!]
\centering
\setlength{\tabcolsep}{5.5pt}
\captionsetup{width=6.8in}
\caption{ Basic information of new GRSs identified from MRC catalogue as described in Sec.\ \ref{sec:newgrgs}. The RA and Dec represent the location of the host galaxy. S$_{\nu}$ is measured integrated flux density of the source at the given frequency $\nu$. Information has been obtained and combined from \cite{McCarthy96} and \cite{kapahi98}. The 408 MHz flux densities from MRC have been converted to the \citet{Baars1997} scale. Hence, both 408 MHz and 4800 MHz flux densities are in the same flux scale.}
\begin{tabular}{lcccccccccc}
\hline
 Sr.No. & Name & RA(J2000) & Dec(J2000) & Type & LAS & $z$ & Size & S$_{408}$ &  S$_{4800}$ &  $\alpha_{408}^{4800}$ \\
 & & (H:M:S)&(D:M:S) & & ($\arcsec$) & & (Mpc) & (mJy)  & (mJy) &  \\ 
\hline
 1 & MRCB0148$-$297 & 01:50:35.85 & -29:31:56.01 & G&138   & 0.41 & 0.77 & 6829	& 923 & 0.81 \\
 2 & MRCB0430$-$235 & 04:32:52.77	& -23:24:28.01 & G&91  & 0.82 & 0.71 & 931  & 113 & 0.86 \\ 
 3 & MRCB2038$-$280 & 20:41:43.77	&-27:52:51.01 & G&150 & 0.39 &0.82 & 1426	&69&		1.23 \\ 
\hline
\end{tabular}
\label{tab:mrc}
\end{table*}


It is relevant to note here the way angular sizes are usually measured to estimate the projected linear sizes.
Traditionally, for FRII type RSs or GRSs, the largest angular size (LAS) of the source is measured from hotspot to hotspot and for FRIs, LAS is measured from the outermost contours which are often at 3 times rms noise level. In the case of sources with multiple hotspots, the most prominent compact hotspot is taken. For source which appear to have a hybrid morphology \citep[HyMORS;][]{saikia96hymorph,GKhymorph,gkhymors02} the LAS of the source is estimated from the hotspot to the outermost contour of the lobe without a hotspot. The projected linear size is estimated from the LAS using the relation:-

\begin{equation}
  \rm  D = \frac{\theta \times D_{c}}{(1+z)} \times \frac{\pi}{10800}
\end{equation}
where $\theta$ is the LAS of the GRS in the sky in units of arcminutes, D$_{\rm c}$ is the comoving distance in Mpc, $z$ is the host galaxy's redshift, and D is the projected linear size of the GRS in Mpc.

\section{Radio properties}
\subsection{Morphology}\label{sec:morph}
The vast majority of GRSs appear to have an edge-brightened FRII structure with some in the intermediate FRI-II category, and only a small fraction have been classified as FRIs. In the early compilation by \citet{Ishwara1999} only 4 of the 53 ($\sim$7.5\%) GRSs have been classified as FRIs. 
\citet{Kuzmicz2018} have classified 20 of the 349 GRSs in their compilation as FRIs, yielding a percentage of $\sim$\,5.7.  In the LoTSS sample of GRSs, 18 of the 239 GRSs ($\sim$7.5\%) are FRIs \citep{PDLOTSS}, while in the SAGAN\footnote{\href{https://sites.google.com/site/anantasakyatta/sagan}{SAGAN: Search and Analysis of GRGs with Associated Nuclei}} sample, 8 of the 162 GRSs ($\sim$\,4.9\%) are FRIs \citep{sagan1}. Approximately 90\% of the sources in both the LoTSS and SAGAN samples are classified as FRIIs \citep{PDLOTSS,sagan1}. Although this may be due to more  dissipative jets in FRI sources; sensitive low-frequency observations would help clarify whether the percentage of FRI GRSs is larger. 

The fraction of sources with a hybrid morphology, where one side appears to have an FRI structure without hotspots while the opposite side exhibits an FRII structure is also extremely small. It is unclear whether these are genuine FRIs as they usually do not exhibit  
jets as seen in FRI sources, and the absence of a prominent hotspot may not be adequate to classify a side as one with an FRI structure \citep{Saikia22}. Moreover, \citet{harwood20} have pointed out that most hybrid morphology sources are intrinsically FRII sources
and their apparent hybrid morphology is the result of orientation effects where the lobes are not parallel to the inner jet. 

Using sensitive radio observations of 8 FRII GRSs, \citet{ravi96} showed that the axial ratio for GRSs is similar to that of smaller radio sources. They also indicated that GRSs tend to have less uniform bridges compared to smaller radio sources; however, GRSs in denser environments (e.g. galaxy cluster centres) showed the presence of more uniform bridges. These trends suggest that the FRII sources undergo self-similar morphological evolution with the nature of the ambient medium playing a significant role. In addition to this, they also report observing more intermittent jet activity in GRSs. We discuss the recurrent jet activity and double-double radio morphology of GRSs in Sec. \ref{sec:rejuv}

The prominence of the bridge emission could also be affected by inverse-Compton scattering with the cosmic microwave background (CMB). \citet{Ishwara1999} showed that inverse-Compton losses dominate over synchrotron losses for the GRSs while the reverse is true for the smaller sources. This would also have an effect on the prominence of the bridge emission with redshift. \citet{Konar2004} investigated this aspect for a small sample of GRSs and found the bridge prominence to decrease with redshift. With the large number of GRSs now available it would be interesting to re-examine this trend.

\subsection{Symmetry parameters}
The symmetry parameters of double-lobed FRII radio sources include the separation or arm-length ratio, R$_\theta$, defined as the ratio of separation of the outer hotspots from the radio core or host optical galaxy; the corresponding flux density ratio, R$_s$, and the misalignment angle, $\Delta$. The latter is defined as the supplement of the angle formed at the nucleus or radio core by the outer hotspots. For an intrinsically symmetric source in a symmetric external environment, R$_\theta$, depends on the light travel time across the source and is given by $(1 + \beta cos\phi)/(1 - \beta cos\phi)$ where $\beta c$ is the velocity of advancement of the hotspots and $\phi$ is the angle of inclination of the source axis to the line of sight. 

One of the early attempts to examine the separation ratio of GRSs and smaller sources was by \citet{Ishwara1999}  who found  the median value of R$_\theta$ for GRGs to be $\sim$\,1.39, marginally higher than that for a sample of 3CR radio galaxies of smaller size where R$_\theta \sim$1.19. \citet{Schoenmakers_spec} carried out multi-frequency studies of 26 GRGs selected from the WENSS, and found a similar result for the GRGs to be marginally more asymmetric. \citet{lara04}, \citet{Komberg09} and \citet{Kuzmicz12} however found no significant difference. Although compact radio galaxies of sub-galactic sizes are found to be more asymmetric due to asymmetries in the interstellar medium of the host galaxies (e.g. \citealt{saikia03}), the existence of asymmetric GRSs suggests the existence of asymmetries in the external environment on Mpc scales as well. In the unified scheme for radio galaxies and quasars, the quasars are expected to be at smaller angles to the line of sight so that orientation effects are more significant than for galaxies. The quasars are indeed found to have a flatter distribution of R$_\theta$ for smaller sources. \citet{Ishwara1999} also pointed out that in the four GRGs and 2 GRQs in their sample with radio jets, the lobe on the jet side is closer in three of the four galaxies but farther in both the quasars. In the GRQ 4C34.47 which also exhibits superluminal motion and has a kpc-scale radio jet, the hotspot facing the jet is farther and more compact and brighter \citep{hocuk10}, consistent with the unified scheme. But this is not always the case. Among the GRQs with jets listed by \citet{kuzmic21}, the hotspot facing the weak jet in J1048+3209 is nearer and brighter. The separation ratio of the jetted side to the opposite side is 0.69 while the corresponding flux density ratio is $\sim$2. Overall this suggests that the environment plays an important role in galaxies, while orientation effects could be dominant for most quasars, consistent with the unified scheme, while there are examples of GRQs with large environmental asymmetries as well. 

The flux density ratio R$_s = ({\rm R}_\theta)^n$ where $n = 2, 3$ depending on the physical situation (e.g. \citealt{Blandford79}). Again in an ideal symmetric source in a symmetric environment, the side facing the jet should be farther from the core and usually brighter. \citet{Schoenmakers_spec} found in their sample of GRGs
that in 13 out of their 20 sources the brighter lobe is on the nearer side, suggesting the importance of environmental effects. 

Assuming that the distribution of galaxies in the vicinity of a GRS reflects the ambient density distribution in the intergalactic medium, a number of authors have examined its effect on the structure of the sources. Several authors (e.g. \citet{saripalli86} for a small sample, \citet{ravi08} for B0503-286, \citet{Safouris09} for B0319-454, \citet{chen11ngc6251} for NGC6251, \citet{chen11da240,chen18da240} for DA240, \citet{chen12ngc315} for NGC315, \citet{chen124c73} for 4C73.08, \citet{malarecki15} for a sample of 19 GRGs) have reported evidence of anisotropies in the medium affecting the structure of the GRSs, with the shorter lobe being on the side with larger galaxy density. 
\citet{malarecki15} also noted that for non-collinear sources the radio lobes are deflected away from regions of high galactic density.
\citet{Pirya12} examined the distributions of galaxies in the environments of 16 large radio sources. They found that for the  GRG J1552+2005 (3C326) which has the highest separation ratio in their sample, there is a group of galaxies which forms part of a filamentary structure in the vicinity of the shorter arm.  Although most of the large sources appear to occur in regions of low galaxy density, \citet{Pirya12} found the shorter arm is brighter in most cases. This suggests that there could be asymmetries in the intergalactic medium which may not be apparent in the distribution of galaxies but may be detectable in deep x-ray observations.

The distributions of misalignment angle for the GRGs and smaller-sized 3CR radio galaxies are similar, with median values of \simi5\deg, suggesting that the smaller-sized RGs and GRGs are basically similar. The quasars in the 3CR sample have a flatter distribution possibly due to projection effects (cf.
\citealt{Ishwara1999}; \citealt{Schoenmakers_spec}). However, these early studies were dominated by galaxies, as the number of GRQs were few in number. From the large samples of  both GRGs and GRQs which have been identified (e.g. \citealt{Kuzmicz2018,PDLOTSS,sagan1,kuzmic21,sagan3}; see Sec.\ \ref{sec:searchmain}), it would be interesting to re-examine these trends.

\subsection{VLBI-scale structure}
The VLBI observations in the radio band allow us to achieve milli-arcsecond (mas) resolutions routinely and also micro-arcsecond resolutions as with the Event Horizon Telescope \citep{EHTmag}. This enables us to study the innermost parts of the radio core and jets, and also the supermassive black hole as in the case of M87. One of the primary requirements for VLBI observations is bright nuclei with flux densities greater than $\sim$\,20 mJy at GHz frequencies. Only a small fraction of GRSs (mostly powered by quasars) are bright enough to be observed with VLBI currently. Hence, only a few GRSs have been studied at VLBI resolutions to probe their core and inner jet structures.  In order to test the unification scheme of quasars and RGs, \citet{Saripalli97} carried out a VLBI survey of 6 GRGs and studied  a sample of 8 GRSs (4C39.04, 3C130, DA240, 4C73.08, HB13, 1245+673, PKS1331$-$09, and 3C326A). Two of these sources (DA240 and PKS1331-09) were previously observed by \cite{Graham81} as part of a bigger sample of bright sources (core flux densities greater than 100 mJy at 4.8 GHz). Under the unification scheme, the GRGs are expected to be oriented between $\sim$45\deg and 90\deg ~to our line of sight, while GRQs are inclined at smaller angles. \citet{Saripalli97} presented their findings based on 25mas resolution maps at 1.67 GHz for the sample, where they find compact cores in all the eight sources and asymmetric core-jet structure in PKS1331-09. They suggest that the cores may not be Doppler beamed based on the overall orientation of the GRSs. They also find that the core strength is similar to that of normal sized RGs. The GRG DA240 in their sample showed pc scale bipolar jets, and hence they infer it to be oriented close to the sky plane.  VLBI observations of the central source in the GRG 3C236 (see Fig.\ \ref{fig:3c326ims}) by \citet{schilizzi19883c236} and \citet{Schilizzi01} show an asymmetric jet from which they estimate an inclination angle of $\sim$30$^\circ$ to the sky plane. Although large angles of inclination to the line of sight for galaxies would appear to be consistent with the unified scheme, the suggested inclination angle for the GRG 7C 1144+3517, which also exhibits superluminal motion is $<$~25\deg \citep{Giovannini99}.

\citet{Barthel89vlbi} and \citet{Hooimeyer92} found superluminal motion in GRQ 4C34.74 and also a projected misalignment angle of 5$\deg$ between the large- and the small-scale jets. Over the years the radio core has been found to be variable as well as the host quasar at optical wavelengths \citep{hocuk10}. Combining VLBI-scale results with the large-scale radio emission observation for this source, \citet{hocuk10} estimated the line of sight angle to be between 53$\deg$ and 57$\deg$. GRQ 4C34.74 shows a strong one-sided jet on pc to kpc scale, extending up to \simi400 kpc, which is one of the largest radio jets observed along with RG CGCG 049-033 \citep{bagchi07}. On VLBI scales, for the jet of GRQ 4C34.74, \citet{Barthel89vlbi} and \citet{Hooimeyer92} report a knot proper-motion of 0.29 mas~year$^{-1}$, which gives the projected expansion speed of 3.9\textit{c}.

Like most of the GRSs or radio sources in general, GRQ 4C74.26 also shows a one-sided jet even on pc scales \citep{pearson92}. However, unlike GRQ 4C34.74, GRQ 4C74.26 does not show misalignment between the large- and small-scale jets \citep{pearson92}. They argue that if Doppler beaming is responsible for the observed asymmetry of the parsec-scale jet then the jet axis must be $<$ 49$\deg$ from the line of sight. 

For GRG 4C+69.15, \citet{liu09} found a steep spectrum core at VLBI scales (\simi2mas) between 2.3 and 8 GHz, indicating signs of restarted activity. Further high-resolution VLBI observations are needed to clarify its structure.

NGC~6251 is a GRG with a size of $\sim$\,2\,Mpc growing in a relatively dense environment, having  a prominent $\sim$\,200\,kpc jet \citep{Waggett77}, also referred to as `blowtorch jet'. Its counterjet of 
$\sim$\,50\,kpc was detected using the VLA \citep{Perley84}. The jet has been studied in detail from radio to x-ray wavelengths \citep[e.g.][]{Perley84,mack97,Evans05}. \citet{DaytonJones86} carried out detailed study of NGC~6251 using VLBI techniques, and observed that the nuclear jet is well collimated and like GRQ 4C34.74, NGC~6251 also shows a misalignment of \simi5$\deg$ between its \simi4 mas inner jet and kpc-scale jet. The large jet to counterjet brightness ratio in NGC~6251 at smaller as well as longer scales can be explained by relativistic beaming. \citet{DaytonJones86} concluded that the overall radio axis of NGC~6251 is oriented $<$45$\deg$ to our line of sight.

 The large-scale structure of NGC315 exhibits a long collimated highly polarised jet which terminates with a sharp bend. It has a faint counterjet, which was not detected at pc scale in the earliest VLBI scale observations as reported in  \citet{Linfield81}. The \simi2~mas resolution VLBI observation of NGC315 by \citet{cotton99} revealed the faint counterjet. \citet{cotton99} by combining their results with that of other VLBI results for NGC315 from the literature (e.g. \citealt{Venturi93}) conclude that the core is in an active phase with occasional increment in the continuum radio flux. \citet{cotton99} also note that in contrast to the highly polarised nature of its large-scale jets, the pc-scale jet is unpolarised at 6~cm and they attribute this to the disorganised magnetic fields at pc scales or Faraday depolarisation occurring due to the narrow-line region (NLR). Recent sensitive VLBI observations by \citet{park21} reveal jet structures down to sub pc scale and find an indication of limb brightening in the jet at 43 GHz. They also constrain the jet viewing angle to \simi52$\deg$, which is in accordance with estimates from kpc scale jets \citep{Laing14}.

\subsection{Spectral Index ($\alpha$)}
Since the GRSs grow to megaparsec scales from smaller sizes, and the basic physical processes are  similar to RSs, it is expected that they may have steeper spectral indices than the smaller sized RSs. \citet{PDLOTSS} using the LoTSS sample and \citet{sagan1} using a large compendium of GRSs, showed that the mean ($\alpha$ $\simeq$0.75) spectral index of GRSs is similar to that of SRSs. However, as there is a significant correlation between spectral index and luminosity (e.g. \citealt{Laing1980}), a more detailed comparison of spectral indices of GRSs and smaller sources in well-defined luminosity bins would be useful.

Spectral index variation along the lobes of FRII radio sources have shown that their spectra steepen with distance from the hotspots (also for GRSs; e.g. \citealt{Schoenmakers_spec}), as expected in the models of radio sources. In one of the early attempts, \citet{leahy89} compared MERLIN images at 151 MHz with VLA images at 1.5 GHz for a sample of 3CR sources and found that spectral indices could increase to $\sim$ 1.4-2 towards the centre. Similar studies have been done for a number of GRSs and used to determine the spectral ages of radio sources as discussed in Sec.\ \ref{sec:specage}.

\subsection{Polarisation and Magnetic field structure}\label{sec:pol}
Polarization observations of RLAGN give us valuable information on the source itself and could also be used to probe the immediate environment as well as the medium through which the radio waves propagate. The observed degree of linear polarization could be affected by how uniform or tangled the magnetic field lines are, and also by Faraday rotation of the  
polarized signal by thermal plasma within the source or along its path. These aspects along with the observed polarization properties of extragalactic radio sources have been reviewed by \citet{1988Saikia_pol}. Although the broad trends discussed in the review have remained similar, there have been detailed studies of individual sources in recent times with MeerKAT (IC4296: \citealt{Condon21IC4296}; PKS2014-55: \citealt{cotton20}; ESO137-006: \citealt{Ramatsoku20}) and at low frequencies with LOFAR \citep{Mahatma21}. Here, we highlight a few studies of GRSs, which tend to have ordered magnetic fields along the lobes with high degrees of polarization at cm wavelengths and low rotation measure. As the GRSs are much larger than the host galaxies, and mostly tend to occur in regions of low galaxy density, they can be valuable probes of the intergalactic medium, including its magnetic field.

Detailed studies of several well-known GRSs were initially done with the WSRT usually at 49, 21 and 6~cm. 
The polarisation study of GRG 3C326 by \citet{Willis78}
showed that the lobes are $\sim$40\% polarized with very little depolarization between 21 and 49~cm from which they estimate the density of thermal material in the lobes to be $\sim$ 2 - $6 \times 10^{-5}$ cm$^{-3}$. The rotation measure (RM) across most of the lobes is within $\sim\pm$2 of $\sim$20 rad m$^{-2}$, and the magnetic field is uniform over several hundred kpc and oriented predominantly along the major axis of the source. Although differing in detail, similar trends were seen in the GRG 3C236 as discussed in \citet{Strom80}. Unlike 3C236 and 3C326, the GRG NGC315 has very prominent jets which were studied by \citet{Willis81NGC315}. 
The dominant magnetic field is initially along the jet axis and changes to a well-ordered orientation perpendicular to the jet axis. Such changes are observed in jets in FRI and FRII sources while those in FRII sources are predominantly along the source axes \citep{Bridle84,Saikia22}.
Studies of several giant sources have shown similar results
\citep[e.g.][]{Tsien82,Kronberg86,Jaegers87,lara00}.

\citet{Machalski06pol} have mapped the magnetic field structure of a sub-sample 17 GRGs using the VLA (4.9 GHz) and reported the degree of polarisation to be in the range of 3-18.5\% for the lobe closer to the core, and 3.5-32.6\% for the lobe  which is farther from the core. The respective RM values are in the range of $-$8 to 18 rad~m$^{-2}$ and $-$8 to 14 rad~m$^{-2}$, respectively. For most of the sources, the hotspots are not well resolved in the radio maps and their degree of polarisation is low. 

Polarization observations at low frequencies would enable us to determine RMs over a large frequency range, and also put useful constraints on thermal material from depolarization and RM values. An early attempt of observing GRSs at low frequencies was by \citet{mack97} who detected polarization at  
326 MHz using the WSRT. The polarization capability of LOFAR has extended polarization studies to lower frequencies and yielded a number of interesting results.  
High-resolution LOFAR imaging of double double giant radio galaxy B1834+620 \citep{2015Orru} at 144 MHz has detected polarised emission at +60 rad~m$^{-2}$ in the northern outer lobe, consistent with earlier RM estimate +55 - 60 rad m$^{-2}$ by \citet{SchoenmakersB1834+620} from higher-frequency observations. \citet{SchoenmakersB1834+620} find no significant evidence of depolarization, and the RM is likely to be largely of Galactic origin.

Using a sub-sample of 179 GRSs from \citet{PDLOTSS}, \citet{Stuardi20} studied linear polarisation, Faraday rotation measure, and depolarisation properties of GRSs. They have used LoTSS (144 MHz) at low frequencies and NVSS (1400 MHz) at higher frequency for this study, where they detected polarisation in 36 GRSs from the LoTSS sample and in 3C236. Owing to their high precision (\simi0.05~rad~m$^{-2}$) RM measurements, they detect considerable RM differences between the lobes of the sources with a median value of \simi1~rad~m$^{-2}$. Considering the detectable polarised parts of the GRSs (e.g. core, hotspots, and lobes), the RM values are in the range of 3 to 28 ~rad~m$^{-2}$ with a median of 12.8~rad~m$^{-2}$. They also observe that the polarisation fraction is higher in larger GRSs. They detect Faraday depolarization caused by a Faraday dispersion of up to $\sim$0.3 rad m$^{-2}$, which has been possible due to the low-frequency observations. They suggest that this may be due to small-scale magnetic field fluctuations in the vicinity of the GRSs. From their observations \citet{Stuardi20} infer that the GRSs are normally in a very low-density environment with thermal 
electron densities $<10^{-5}$ cm$^{-3}$ and magnetic fields
less than about 0.1$\muup$G.

\subsection{Minimum energy densities and magnetic field strength}\label{sec:uminmag}
In order to understand the physical processes in radio sources, their energetics, propagation and confinement, it is essential to determine the relevant physical parameters. Two of the fundamental parameters are the energy content of the different components and the corresponding magnetic field strengths. Traditionally the minimum energy condition, in which there is approximate equipartition between the particle energy density and the magnetic energy density has been used to estimate these two quantities. The required inputs are the source luminosity, volume and spectral index which can be reasonably well determined, and the relatively unknown ratio of total energies of cosmic ray protons and electrons. \citet{Beck_Krause} have presented revised formulae incorporating the total energy of cosmic ray nuclei to that of the synchrotron emitting electrons and positrons.
A more direct estimate of the magnetic field strength can be made from the observations of high energy x-rays from radio sources or their components, where the high-energy emission is due to inverse-Compton scattering of ambient photons by the low-energy electrons. The photon field could be the cosmic microwave background (CMB) radiation, synchrotron photons, or ambient photons from starlight, AGN or the extragalactic background. The x-ray emission gives an estimate of the inverse-Compton scattering electrons. From the observed synchrotron emission, the magnetic field strength can be estimated. For more detailed discussions, see \citet{Pacholczyk}, \citet{Miley}, and \citet{HardcastleCrostron20}.

In the last \simi40 years, various studies have estimated the minimum energy densities (u$_{\rm min}$) of GRSs, where the number of estimates from the radio method is far larger compared to the estimates from x-ray method. It is relevant to compare these estimates with one another and also with the SRSs.
\citet{ravi96} found median u$_{\rm min}$ of \simi8$\times$10$^{-13}$ erg~cm$^{-3}$ for eight luminous GRGs. These estimates of u$_{\rm min}$ for GRGs were found to be quite low when compared to the RSs sample (\simi5$\times$10$^{-12}$ erg~cm$^{-3}$) of \citet{lw84}. \citet{Ishwara1999} found similar median value of u$_{\rm min}$\simi4.4$\times$10$^{-13}$ erg~cm$^{-3}$, with the help of a larger GRS sample. Similarly, \citet{Schoenmakers_spec} with the WENSS sample of 20 GRSs obtained median u$_{\rm min}$\simi4.8$\times$10$^{-13}$ erg~cm$^{-3}$. \cite{Konar2004} using a sample of 17 GRSs found a slightly higher value of u$_{\rm min}$\simi 11.9$\times$10$^{-13}$ erg~cm$^{-3}$. 
Also, \citet{M04} compared u$_{\rm min}$ values of GRSs with RSs and found these to be less than those of RSs. 

\begin{figure}
    \includegraphics[scale=0.32]{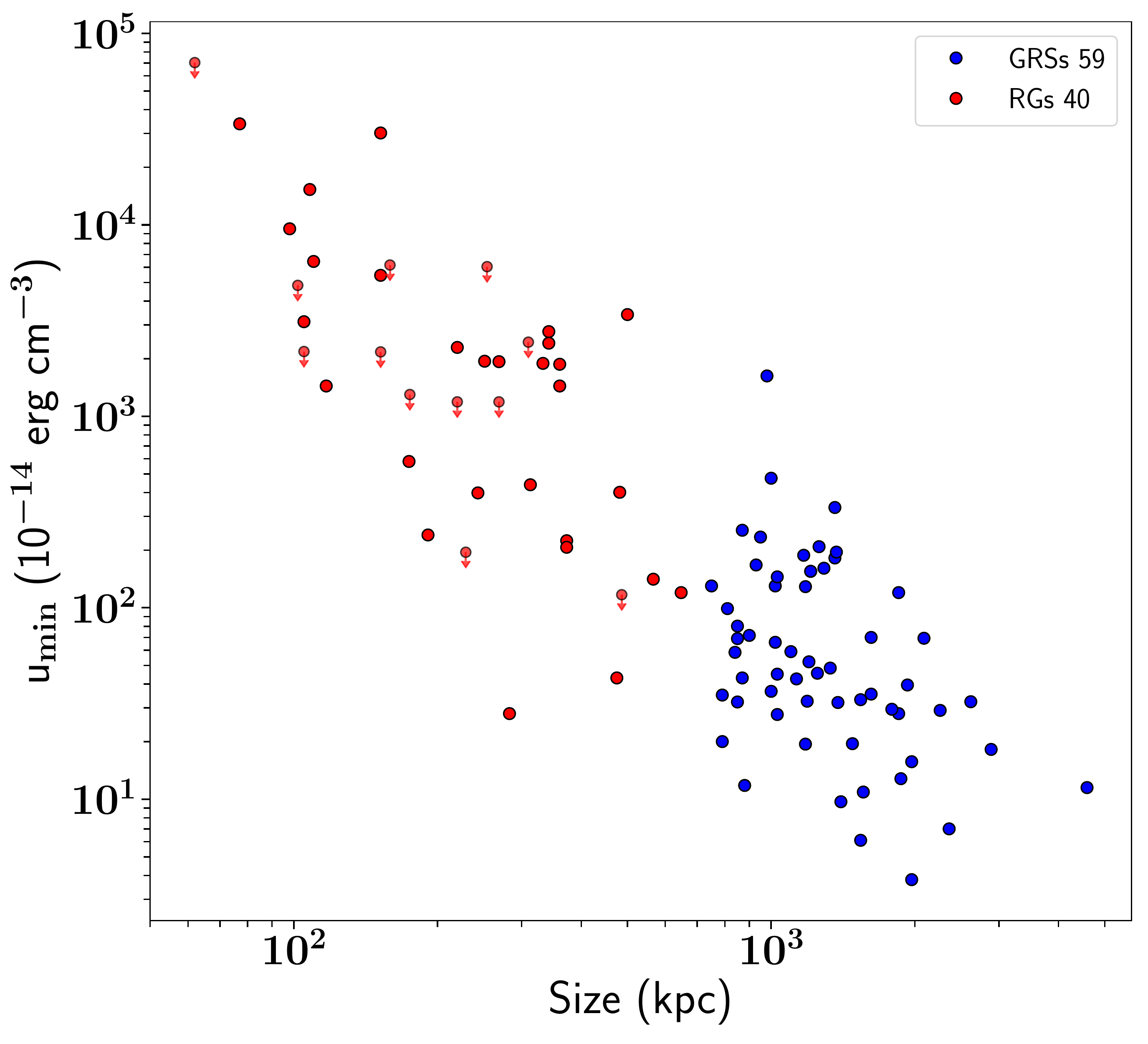}
    \caption{Electron minimum energy densities of FRII RGs and GRSs as a function of their projected linear sizes as discussed in Sec.\ \ref{sec:uminmag}.}
    \label{fig:uminsize}
\end{figure}

Using x-ray data to detect inverse-Compton x-rays from the lobes of GRSs, it is possible to estimate the electron energy density (u$_{\rm e}$), magnetic energy density (u$_{\rm m}$), and magnetic field (B), without assuming equipartition condition. \citet{Isobe15} showed (Figure 9 of their paper) u$_{\rm e}$ to be considerably low for GRGs in comparison with RGs. They infer that jet activity in GRGs is already on decline and the energy input to the lobes is insufficient to overcome the radiative and adiabatic losses.
The estimates of four GRGs, namely 3C326 \citep{ Isobe09}, 3C236 \citep{Isobe15}, DA240 \citep{Isobe11DA240}, and 3C35 \citep{Isobe11} were made using the high energy instrument on \textit{Suzaku} space telescope. They have estimated u$_{\rm e}$ for FRII type RGs based on the data from \citet{Croston05}. 
As seen in Fig.\ \ref{fig:uminsize} we have plotted u$_{\rm min}$ (\simi u$_{\rm e}$) as a function of projected linear size, where we have used data from \cite{Isobe15} for RGs and GRGs, along with estimates (using radio observations) for GRGs from a number of papers \citep{ravi96,Ishwara1999,Konar2004,Schoenmakers_spec}. We observe a clear trend of u$_{\rm min}$ decreasing as the size of the source increases. 
The SKA with its unprecedented sensitivities, will be able to find GRGs with much lower u$_{\rm min}$ and help us understand the physical processes involved here better.

The estimates of the magnetic field strengths (B$_{\rm eq}$) from equipartition conditions derived from radio data for GRSs are in the range of B$_{\rm eq}$ \simi1 - 16~$\muup$G with a median value of \simi5~$\muup$G \citep{mack98,Schoenmakers_spec,Ishwara1999,Konar2004,konar06,Machalski06pol,Jamrozy08,Tamhane2015,binny18}. The radio lobes of a few GRSs have been detected in the x-rays using mainly the \textit{Suzaku}, which allows to compute magnetic field strength independently without assuming equipartition condition \citep{harris79}. These estimates of magnetic field (B) are slightly lower than the above mentioned range (3C326: 0.55~$\muup$G, \citealt{Isobe09}; DA240: 0.87$\muup$G, \citealt{Isobe11DA240}; 3C35: 0.88~$\muup$G, \cite{Isobe11}; 3C236: 0.48~$\muup$G, \citealt{Isobe15}) but have been found to be similar to their respective B$_{\rm eq}$. \citet{konar09} studied the GRG 3C457 at radio and x-ray wavelengths using the GMRT, VLA and \textit{XMM-Newton} and found the magnetic field of the northern and southern components from inverse-Compton scattering model to be 6.8 and 4.0 ~$\muup$G respectively. These values are within a factor of 2 lower than the equipartition values.  Similarly, using \textit{XMM-Newton}, \citet{Tamhane2015} and  \cite{Mirakhor21} detected lobes in x-rays for GRG J021659$-$044920 and GRG J2345$-$0449, respectively. While B for GRG J2345-0449 was found to be in the range of 0.22-0.39~$\muup$G for its northern and southern lobes, the value for GRG J021659-044920 was found to be slightly higher with B\simi3.3~$\muup$G, which is interestingly located at a redshift of 1.3.

\subsection{Jet kinetic power ($Q_{Jet}$) }
Jets are the twin collimated output of AGN which transport bulk kinetic energy as these traverse outwards initially through the interstellar medium of the host galaxy and later through the intracluster and intergalactic medium. Either the accretion of gas onto the supermassive black hole or extraction of energy from the spinning black hole or the combination of both is responsible for the launching of jets. The jets start with relativistic speeds at the launching site and gradually appear to decelerate while interacting with the intergalactic medium. In some radio galaxies, the jets end up into diffuse plasma and give rise to FRI type morphology. For more powerful radio galaxies, the highly collimated jets interact with the intergalactic medium to form regions of bright shocks called hotspots. They are classified as FRII radio galaxies. An important estimate of jet kinetic power (\qj) is the x-ray cavity power produced by the lobes in the external medium. The amount of work done to excavate such observed x-ray bubbles provides a good measurement for jet kinetic energy of AGN. The method requires the knowledge of age of the radio galaxy and the work done p$\Delta$V from x-ray spectroscopy \citep{birzan04}. It can be achieved for a small population of sources, especially in rich clusters whose x-ray measurements of cavities are possible. Also, the parameter of age is not well constrained for radio galaxies. Due to these biases and uncertainties in this process, extreme caution should be taken. \citet{Schoenmakers_spec} have reported the jet power of 7 GRSs with \qj value nearly $\sim$10$^{44}$ erg~s$^{-1}$. They have determined \qj by dividing the total energy contents of the lobes by the age of the source. For a hard x-ray selected sample of 14 GRSs, \citet{Ursini18xgrg}, find $\rm Q_{Jet}$ in the range of $\sim$\,10$^{42}$ erg s$^{-1}$ to 10$^{44}$ erg s$^{-1}$. In the absence of the knowledge of age, internal magnetic field strength and other related parameters, \citet{sagan1} and \citet{sagan3} have adopted the method based on simulation-based analytical model of \citet{Qjet_Hardcastle} which involves determining jet power by probing synchrotron emission through low-frequency radio observations. However, high-frequency radio observations are ideal to explore jet components as they have flat spectral nature. But at higher frequencies, Doppler boosting effect is very prominent. Therefore, low radio frequency observations are preferred to probe \qj in order to counteract Doppler enhancement effects. The relation between low radio frequency luminosity and $\rm Q_{Jet}$ is given as -

\begin{equation}
\rm  L_{150} = 3 \times 10^{27}\frac{Q_{\rm Jet}}{10^{38} \ W} W \ Hz^{-1}
   \label{jet_kp2}
\end{equation} 

Using the above relation, the jet kinetic power of GRSs as shown by \citet{sagan1} lie in the range of 10$^{41}$ to 10$^{45}$ erg~s$^{-1}$.
However, when compared with SRGs, \citet{mingo14} have shown that GRGs carry less kinetic energy via jets. This is possibly due to severe radiative losses during the period of growth of GRGs \citep{sagan1}. A study on giant radio quasars (GRQs) by \citet{sagan3} revealed a similar trend in jet power of GRQs (205) when compared with redshift matched samples of small radio quasars (SRQs, 379)  i.e., GRQ jets carry almost half of the radio power as compared to jets of SRQs. For the majority of the sources in all the samples, radio luminosity from the TGSS at 150 MHz has been used to estimate Q$_{\rm Jet}$. And for the GRGs and GRQs taken from \citet{PDLOTSS}, 144 MHz radio luminosity measurements from the LoTSS have been considered. Nevertheless, analytical model of \citet{Qjet_Hardcastle} shows a scatter of 0.4 dex (rms) for the sources below redshift 0.5. The model provides good results for FRII sources, however, the measurements are underestimated in the case of diffuse giants and remnants \citep{Qjet_Hardcastle}. The environment and source age have a significant effect on the evolution of radio luminosity as well as jet power. The adjustable parameters should be refined and experimented on various AGN samples for universally robust outcomes. Recently, \citet{Machalski21} have modelled the dynamical evolution of 361 FRII radio sources using DYNAGE algorithm and its extension KDA\footnote{Kaiser, Dennett-Thorpe \& Alexander; \citealt{KDA97}} EXT model \citep{Kuligowska2017}, and found their jet power to lie in the range of 10$^{41}$ to 10$^{47}$ erg~s$^{-1}$. On analysing the sources from their sample, we have found that in the redshift-matched bin of 0.03 $\leq$ $z$ $<$ 0.9, SRGs (153) and GRGs (41) have similar distributions of jet power. The jet power distributions of SRQs (30) and GRQs (6) have also been found to be similar for the redshift-matched (0.5 $\leq$ $z$ $<$ 1.2) samples. 
It would be interesting to extend this analysis to a larger number of sources.

\subsection{Spectral and dynamical ages}\label{sec:specage}
Observations have shown gradual steepening of radio spectra from hotspots towards the core in radio sources, which supports the theory of radiative ageing of relativistic electrons, also known as spectral ageing. The spectral age or the radiative age refers to the time which has passed since the electrons in parts of the radio source were last accelerated. The rapid acceleration of electrons is most likely to occur in the knots of the jets and the hotspots, and these electrons grow older or age as they diffuse away from the hotspots. In the case of hotspots, the ageing electrons form the backflow as part of the lobes, which often extend quite close to the radio core. The resultant observed steepening of the radio spectrum of these parts of the radio source owing to the radiative losses can provide an estimate of the spectral ages.
A radiating population whose energy distribution is initially a power-law, after suffering  synchrotron losses, shows a `break' in the power-law at later times. The frequency at which the break occurs is known as the break frequency ($\nu_{br}$) and it shifts to lower frequencies as
more time elapses \citep{Karda62,Pacholczyk,JP73} predicting steeper measured spectral indices for the older emission.
The expansion of the sources is in good agreement with this. However, the above is not caveat free and the details can be found in \citet{Blundell00} and \citet{brats13}.
The connection between spectral age and radio spectrum gives us insight into the growth and evolution of radio sources over cosmic time and this requires careful analysis of the radio spectra across the source dimensions. High frequency steepening occurs over a broad range of radio frequencies. Thus proper analysis requires sensitive observations over a broad range in frequencies  of matched spatial resolutions, as the spectrum is not same throughout the extent of the radio source.

Spectral ageing studies of GRSs can tell us how long the sources have been active and test the third model/suggested explanation given in Sec.\ \ref{sec:models}. It requires a broad range of  multi-frequency data to study the overall spectral nature of the sources and find the spectral breaks to estimate the spectral ages  and the magnetic field strengths.
To estimate the spectral age in different regions of the sources first we have to calculate the magnetic field in the corresponding regions. The equipartition energy density (U$_{\rm min}$) and the corresponding magnetic fields (B$\rm _{eq}(cl)$) are calculated by employing  classical formalism \citep{Miley,Govoni04}:

\begin{equation}
\rm B_{eq}(cl)= \left({{24\pi}\over{7}} U_{min}\right)^{1/2},
\label{eq:Beq}
\end{equation}

\noindent The spectral or radiative age (t$\rm_s$ ; Myr) can be calculated from  break frequency ($\nu_{br}$ ; GHz) and magnetic field B or B$\rm _{eq}$~($\muup$G) by using the following formula \citep{vanderLaan1969}:

\begin{equation}
\rm t_s=1590\frac{B^{\rm 0.5}}{(B^2+B_{\rm CMB}^2)\sqrt{\nu_{\rm br}(1+z)}} \\ ,
\label{eqtime}
\end{equation}

\noindent where B$\rm _{CMB}=3.18\cdot(1+z)^2$~$\muup$G is the magnetic field strength equivalent to the cosmic microwave background radiation and $z$ is the redshift of the source. In order to address the uncertainty in the ratio (\textit{K}) of total energies of cosmic ray protons and electrons used in the estimation of the classical equipartition or minimum-energy (also discussed briefly in Sec.\ \ref{sec:uminmag}), \citet{Govoni04} and \citet{Beck_Krause} have provided a revised\footnote{However, most of the estimation of GRS as well normal-sized radio sources was either carried before (e.g. see \citealt{mack98}) this or have not used it.} formalism, which is now commonly used.

In Tab.\ \ref{tab:specage}, we have provided a summary of all major spectral ageing studies of radio galaxies and quasars carried out in the past four decades. The earlier estimates of radio sources including giants were mostly based on GHz frequencies over relatively short-ranges from VLA and WSRT. They lacked essential low-frequency observations ($<$ 200 MHz), which are crucial in determining the injection index and for some cases the $\nu_{br}$. Also, while examining spectral index variations using interferometric observations, it is important to match the uv-coverage when using different telescopes/configurations for the different frequencies.

\citet{mack98} provides reliable estimates of spectral ages of 5 GRSs owing to their wide frequency coverage and usage of single dish 100m Effelsberg telescope observation for higher frequency to counter the missing flux density problem. In order to carry out these spatially resolved estimates of t$\rm_s$, they selected GRGs with very large angular sizes. \citet{mack98} also suggested that the relatively younger ages for GRGs in their sample can be due to the occurrence of re-acceleration in the source.

Based on the compilation given in Tab.\ \ref{tab:specage}, we observe that most normal size radio sources have $\rm t_{s} <$ 70~Myr. However, a few GRSs have t$\rm_s$~$>$ 150 Myr, indicating older source age or older permeating plasma.
As expected from the growth and evolution models for radio sources, it is interesting to note that the age for the sources appears to be longer for larger radio sources as shown by \citet{parma02FR1SPECAGE} and \citet{konar09}.

\begin{figure}
    \hspace{-0.5cm}
    \includegraphics[scale=0.32]{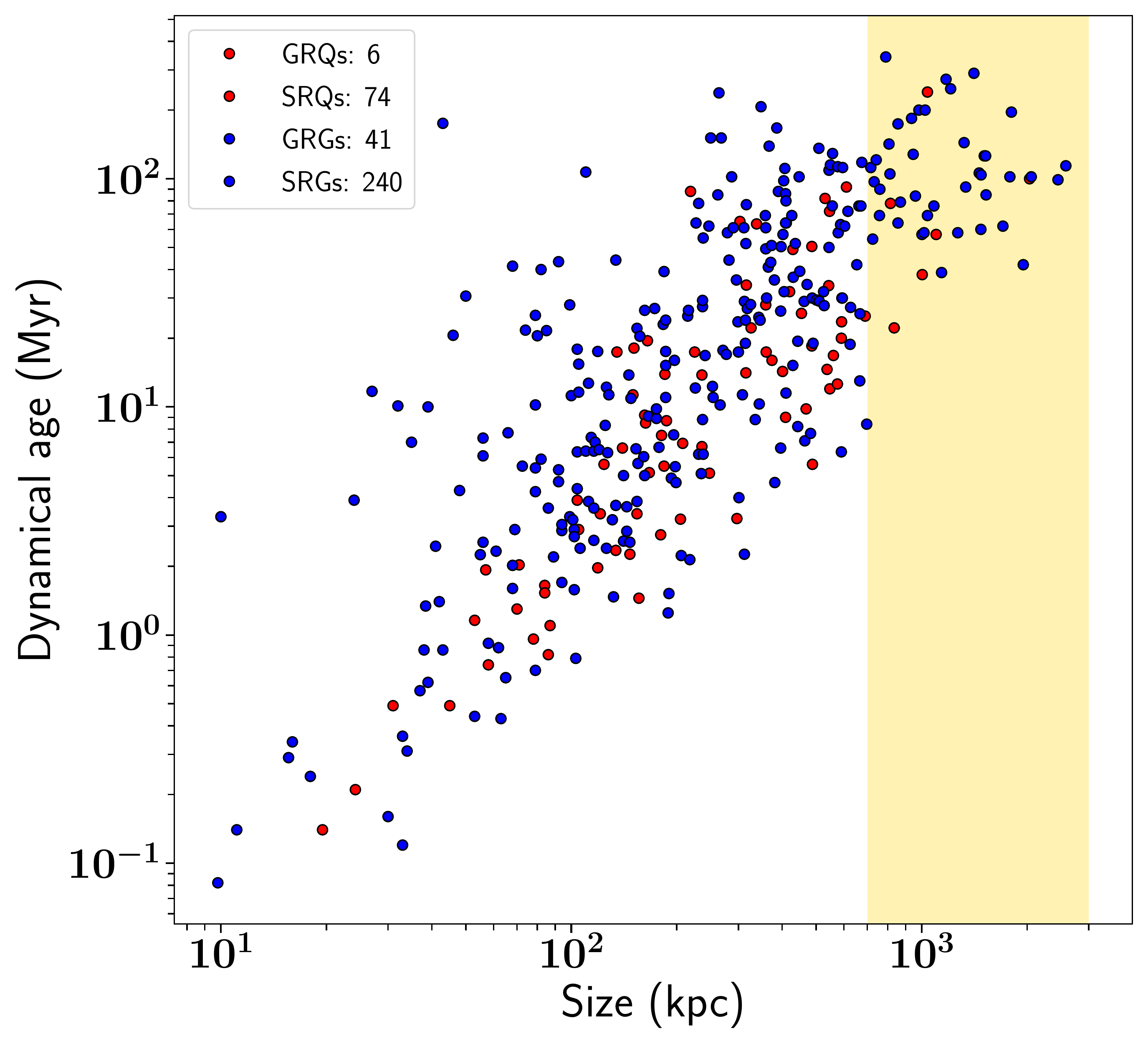}
    \caption{Dynamical age of the radio lobes as a function of the linear size of FRII sources. Data taken from \citet{Machalski21}, where they have used H$_0$ = 71.0 km s$^{-1}$ Mpc$^{-1}$ and $\rm \Omega_m$ = 0.27 as their cosmological parameters to compute the quoted quantities. The shaded region depicts the region covered by GRSs.}
    \label{fig:dagesize}
\end{figure}
While the spectral age estimates in different components of a radio source provide crucial information regarding the past phases, it is equally important to examine the dynamical evolution of the source to understand its actual age. Basically, dynamical age is the estimate of the time elapsed from the birth of the source to the present i.e, the observing epoch. As per the `standard model' of the dynamical evolution of double radio sources \citep{Blanford_Rees74,Scheuer74}, the linear sizes depend on the thrust of the jet or \qj and the ram pressure exerted by the IGM into which the radio source is growing. The internal pressure in the lobes fed by the jet and its interaction with the ambient environment determine the lateral growth or width. This formed the base for various analytical models concerning the properties of radio emission and dynamics of double radio sources \citep{Kaiser97,Blundell99,2002Manolakou}. \citet{2007Machalski} have used KDA model based on the approach of \citet{Kaiser97} to explore source dynamics. An advantage of this model is that the model and its extension can be run using the DYNAGE code \citep{2007Machalski} for the evolution of the radio lobes shortly after the termination of the jet activity. \citet{machalski09} have estimated dynamical ages of the lobes of 10 GRSs using their DYNAGE algorithm and compared them with synchrotron ages \citep{Jamrozy08} of the same sources determined using SYNAGE \citep{Murgia1999}. They found the dynamical ages of the lobes to be 1-5 times more than their spectral ages. This discrepancy is due to the fact that DYNAGE can take into account the radiative effects at lower frequencies more efficiently as compared to SYNAGE \citep{machalski09}. Similar kind of discrepancy between the spectral and dynamical ages were found by \citep{parma99} for low luminosity radio galaxies. According to their results, the dynamical ages are 2-4 times larger than the spectral ages. 
From the recent study of dynamical evolution of FRII radio sources \citep{Machalski21}, it is observed that the range of dynamical ages of the lobes of SRGs and SRQs is 0.082 Myr to 92 Myr, and for GRGs and GRQs it is 38 Myr to 240 Myr. We have taken data from \citet{Machalski21} and investigated if there exists any relationship between dynamical age of radio lobes with linear size of the sources. A linear trend of increase in dynamical age with source size has been found as shown in Fig.\ \ref{fig:dagesize}. It is observed that there is a significant overlap between the spectral ages of RGs and RQs. However, it is quite clear from the plot the giant populations are older ones, although their statistics are comparatively very low.

\begin{table*}[ht!]
\centering
\setlength{\tabcolsep}{1.2pt}
\caption{A summary of spectral age estimates of radio sources from literature. The table is divided into three parts separated by horizontal lines. Here, the top half shows non-GRS samples, the middle shows GRS samples, and the lower part shows individual GRSs. The sequence in each part is ordered as per the frequency coverage used for spectral age estimates. The $^{\dagger}$ represents five double-double GRSs spectral age compilation from \cite{Marecki2021} and references therein.}
\begin{tabular}{lcccccr}
\hline
 Sample&Name & Freq-range & Size  & Age Range & Median Age & Ref  \\
Type $\&$ Size    &     &    (MHz) & (Mpc)   & (Myr)     & (Myr)      &        \\
\hline
14 RSs (3CR) &- & 1400-15000 &0.07 - 0.28 &0.25 - 5.3 & - & \citet{Liu92}\\
38 RSs (MRC) & -&1400-5000 & 0.26 - 0.63 & 2.5 - 94 & 26&\citet{ishwarsaikia01mrc}\\
32 RGs (B2) &- & 1500-5000 & 0.019 - 0.64& 4 - 75 & 28& \citet{parma99}\\
17 RSs (3CR) &- & 151-1500 & 0.14 - 0.60& 2.2 - 145 & 12 & \citet{leahy89}\\
9 RGs (B2) & -    &   600-1400  &   0.19 - 0.65     &   27 - 90        & 55          &  \citet{Klein95}\\

\hline

5 GRGs &-          & 325-10500 & 1.18 - 4.59& 7.6 - 42    & 12 & \cite{mack98}\\
5 GRSs$^{\dagger}$ &- & 70-10000 & 0.9 - 1.4 & 1 - 245    & 15.6 & \cite{Marecki2021}\\
3 GRGs&-       & 240-4800 &1.21 - 1.56 & 5 - 151  & 64& \citet{Godambe09}\\
2 GRGs&-       & 240-4800 &0.85 -1.32 & 30 - 70  & 50 & \citet{pirya11}\\
6 GRGs&-       & 325-4850 &0.84 - 1.79 & 23 - 80 &48.5 &  \citet{Schoenmakers_spec}\\
10 GRS&-       & 325-4800 &0.84 -2.28 & 6 - 46  & 23 & \cite{Jamrozy08}\\
17 GRS&-       & 600-4800 &0.77 - 2.45 & 23 - 45 & 40   & \cite{Konar2004}\\
4 GRS& -&1400-5000 &0.71 - 0.98 & 14.3 - 134 & 34.5&\citet{ishwarsaikia01mrc}\\

\hline
GRG&J0028+0035      & 70-10000 & 1.13& 3.6 - 245 &- &  \citet{Marecki2021}\\

GRG&4C39.04      & 1000-10600 & 1.21& 27 - 62 &- &  \citet{Klein95}\\
GRG&J1453+3308   &   240 to 4860 & 1.37 & 2 - 58  &- &  \cite{konar06} \\
GRG&3C46          & 240-8000 & 0.95 &  2 - 9  &-   &  \cite{nandi10}\\
GRG&8C0821+695   & 150-10600 & 2.62 & 42       &-  &  \citet{lara00} \\
GRG&3C223          & 118-1477 & 0.76 &  84.9 & -  &  \cite{harwood17}\\
GRG&J084408.8+462744 & 325-1400  &2.21 & 13-20 &- & \citet{binny18} \\
GRG& J021659-044920  & 325-1400 &1.24 & 8   & - & \citet{Tamhane2015} \\ 
GRG&3C236          & 143-609 &4.59 &  75 - 159  & -  &  \cite{Shulevski19}\\
\hline
\end{tabular}
\label{tab:specage}
\end{table*}


\begin{figure}[ht!]
    \hspace{-0.5cm}
    \includegraphics[scale=0.5]{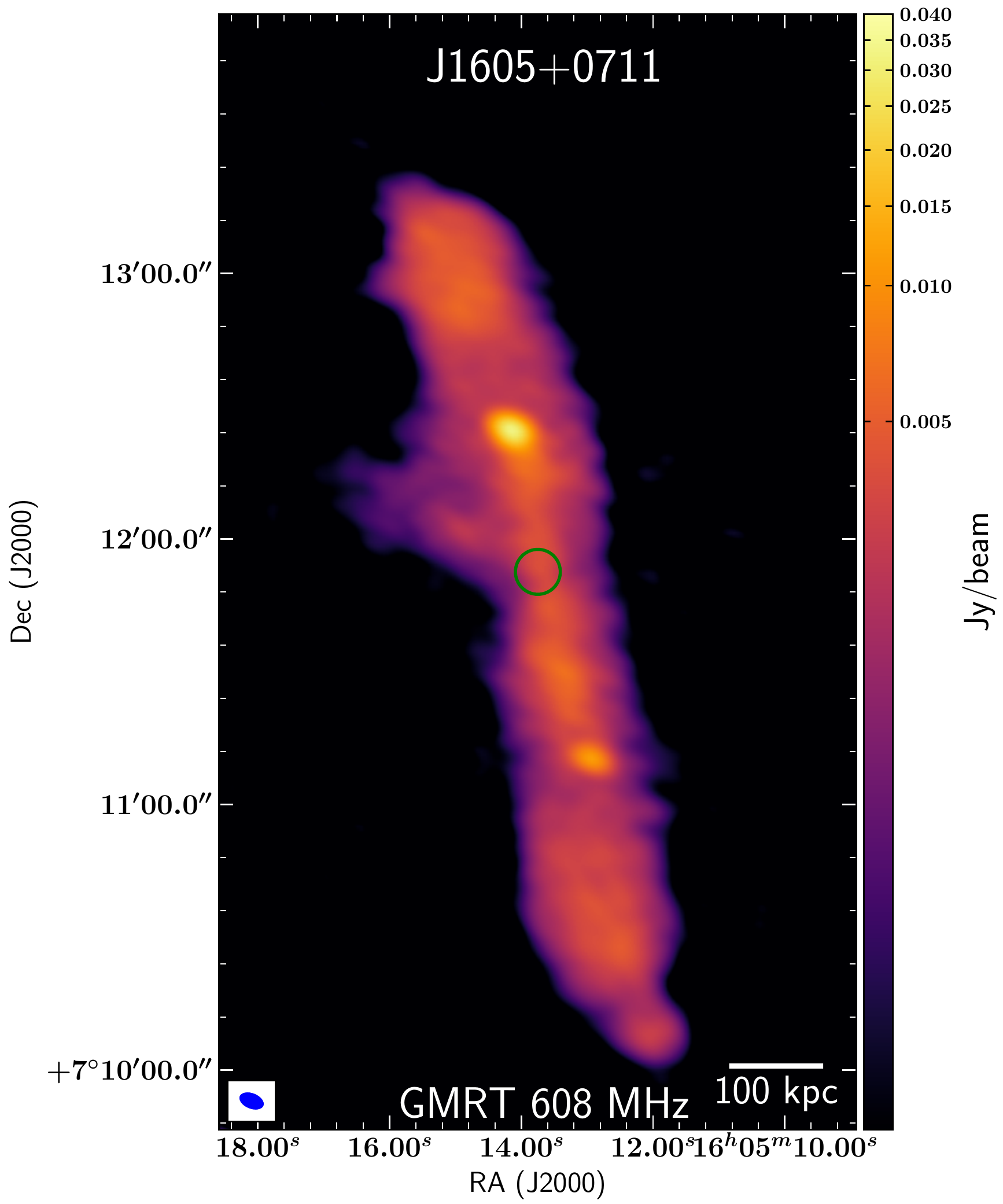}
    \caption{J1605+0711 is an example of a GRG with `double-double' radio morphology, which is an indicator of its restarted activity. Here, we show the GMRT 608 MHz image with the resolution of 6.5\arcsec $\times$ 4.2\arcsec; 68.3$\deg$, represented in blue colour inside white box and rms ($\sigma$) of of 45$\muup$Jy~beam$^{-1}$. The location of the host galaxy is marked with a green circle. A radio core at this location is seen in the VLASS survey too. The radio emission seen is above 3$\sigma$. The overall projected linear size is 874 kpc and the inner double measures about 360 kpc. This source has been studied by \citet{nandi19}, where they estimate its integrated spectral index of $\alpha^{1400}_{608}$ \simi1.41.}
    \label{fig:GDDRG}
\end{figure}

\section{Rejuvenated radio sources}\label{sec:rejuv}

One of the early examples of recurrent nuclear activity was noticed in the radio galaxy 3C338 where a radio jet seen south of the nucleus was interpreted to be due to an earlier cycle of activity \citep{burns83}. Although evidence of recurrent activity may be inferred from sharp discontinuities in the spectral indices in the lobes or from comparisons of radio and x-ray images (see \citealt{Saikia09} for a review); one of the most striking examples of recurrent activity are the double-double radio galaxies (DDRGs;
\citealt{Schoenmakersddrgs1}). These sources have two distinct pairs of radio lobes from two cycles of nuclear activity, the more distant pair being due to an earlier cycle of activity. Many of the early examples of DDRGs found in the northern hemisphere \citep{Schoenmakersddrgs1} as well as in the southern hemisphere \citep{Saripalli02} are associated with GRGs,
where the separation of the outer lobes from the earlier cycle of activity is $>$ 0.7 Mpc (an example is presented in Fig.\ \ref{fig:GDDRG}). \citet{Kaiserddrgs2} suggested a model for these DDRGs where clouds from the external environment are dispersed in the cocoon of the GRGs, facilitating the formation of the second pair of lobes. However, searches for DDRGs among smaller sources also yielded several examples (cf. \citealt{nandi12,nandi19}), and there is also evidence of recurrent activity in compact steep-spectrum and peaked-spectrum radio sources (cf.~\citealt{Odeasaikia21} for a review).
The model developed by \citet{Kaiserddrgs2} is unlikely to work for small sources, and one needs to explore different models for understanding recurrent activity on different size and time scales which have been suggested to range from 10$^4-10^5$ yr to 10$^7-10^8$ yr \citep{Saikia09,Saikia22}.
The variety of signatures of recurrent activity in AGN have also been explored from LOFAR observations at low frequencies which are ideal for detecting diffuse emissions from earlier cycles of activity \citep{mahatma19,Jurlin20,Shabala20}.
\citet{kuzmic17} compiled a sample of 74 DDRGs to study their global properties of which 20 are $>$ 0.7 Mpc. Since the compilation by \citet{kuzmic17}, the number of DDRGs in GRSs has increased, the total number at present being 39 \citep{PDLOTSS,sagan1}. 
Observational evidences suggests that only a small fraction of GRSs are known to exhibit evidence of recurrent activity. For example, in the SAGAN \citep{sagan1} and LoTSS \citep{PDLOTSS} samples of GRSs, where we have carefully examined the structures, 5.1\% and 5.8\% of the sources are DDRGs, respectively. The fraction of DDRGs in the general population of radio sources is also smaller. In the Lockman Hole region, \citet{Jurlin20} suggest that $\sim$\,15\% are restarted sources, but only 3 of the 158 sources with an angular size larger than 60 arcsec are DDRGs. A search for DDRGs in the LoTSS DR1 catalogue revealed only 33 promising candidates from a total of 318542 individual radio sources \citep{mahatma19}. A search for relic lobes around 374 radio sources using deep multifrequency observations with the GMRT revealed no promising candidates \citep{Sirothia2009}.

Using a small sample of hard x-ray selected GRGs and their x-ray properties, \citet{Bassani21} have argued that the AGN of these GRGs in their current phase are accreting efficiently and based on their low lobe luminosities compared to nuclear luminosities, they are restarted AGN.
It is relevant to note here that only about 5\% of GRSs appear to show evidence of recurrent activity in samples such as LoTSS and SAGAN where we have carefully examined the radio structure. Also,  based on the currently available radio and optical data, there are no giant `double-double' radio sources powered by a quasar and they are also quite rare compared to the smaller sized `double-double' radio sources.

One of the most studied GRGs at multiple wavelengths is 3C236, which on VLBI or mas scale shows a double-double morphology as shown in Fig.~\ref{fig:3c326ims}, based on the work of \citet{Schilizzi01}. Another example is the GRG J1247+6723, the bright core with a peaked spectrum is resolved into a compact double on VLBI scales \citep{Marecki03,Bondi04}.  Hence, it is possible that some GRGs or RGs with bright unresolved radio cores (e.g. gigahertz-peaked spectrum core or GPS) may be harbouring an unresolved inner double.

For RLAGN to be rejuvenated one of the ingredients would have to be the availability of fuel, and there have been searches for atomic and molecular gas. Among the GRGs exhibiting a DDRG morphology, both 3C236 and J1247+6723
have been detected in neutral atomic hydrogen \citep{Conway00,Saikia07}. Also, see Sec.\ \ref{sec:costudies} for CO detection in restarted giants.
\citet{chandola10} suggest a higher incidence of H{\sc i} towards the nuclear regions of rejuvenated radio sources for a small sample. This needs further investigation using a large sample of sources.


\section{Host galaxy and AGN properties}
\subsection{Optical}

Like SRGs, almost all the GRGs appear to be hosted by elliptical galaxies with the exception of Speca and J2345$-$0449 \citep{hotaspeca,bagchi14}. \citet{PDLOTSS} from their sample of 239 GRSs, found that at least 40 sources are hosted by quasars. They also concluded that none are hosted by galaxies with spiral morphology based on optical data from the SDSS and Pan-STARRS along with  
Galaxy Zoo classification of \citet{Hart16}.
However, a thorough inspection of deep optical/infrared images of the hosts of GRGs is needed to understand if there exists any oddities in their morphologies as recent deep observations of distant ($z>$ 0.4) galaxies have revealed peculiar shell structures around the galaxies.

\citet{kuzmic19} studied stellar populations of host galaxies for a sample of 41 GRGs (along with surrounding galaxies) and a sample 217 SRGs. They found that the sample of GRGs have more intermediate-age stellar population ($9 \times 10^{8}$ yrs $< \rm t_{*} < 7.5 \times 10^{9} $ yrs) in comparison with SRGs. They also do not find any difference in the mean values of stellar masses of GRGs and SRGs, which is consistent with the findings of \citet{Sabater19}, which is in the context of radio source population in general. Based on their results, \citet{kuzmic19} suggest that it could be a result of past merger activity or cold gas streams permeating through the groups of galaxies in and around the GRG, possibly triggering star formation. 
Recently, \citet{Zovaro22} studied the host galaxy (ESO~422-G028) of GRG 0503-286 in detail using optical integral field spectroscopy from the Wide-Field Spectrograph (WiFeS). They found the host galaxy to have a very old as well as new stellar population and signs of restarted activity. They suggest the possible restarted jet activity and starburst phase could be triggered due to an inflow of gas from a gas-rich merger event.

\citet{kuzmic21} studied samples of GRQs and SRQs using optical SDSS spectroscopic data and found evidence of a correlation between the [O{\sc{iii}}] emission line luminosity and radio power (core and total) for radio quasars. This can be interpreted as a coupling between the radio jet emission and the NLR, or a connection between the powers of photoionising continuum and the radio jet. \citet{KJS2021} constructed a composite optical spectrum for 216 GRQs using SDSS spectra, and found the power-law spectral slope for GRQs to be flatter than that for SRQs and a large sample of SDSS quasars. They also find the quasar continuum to be steeper for higher core and total radio luminosities at 1.4 GHz, but flatter for larger projected linear sizes. They suggest that the slope is orientation dependent, consistent with the unified scheme.

\citet{sagan1} showed for GRGs the absolute \textit{r}-band magnitude (M$_{\rm r}$) lies in the range of $\sim$\,$-$19 to $-$25 in brightness. They find that this distribution of M$_{\rm r}$) for GRGs is similar to that of host galaxies of normal-sized RGs.

Hence, from the above, it appears that  there is not a very clear distinction between the optical properties of GRSs and smaller radio sources.
Upcoming deep wide-field surveys at optical and infrared wavelengths  (e.g. \textit{James Webb Space Telescope} \citep[JWST;][]{Gardner2006,Rigby2022}, Vera C. Rubin Observatory; \citealt{Lsst} and \textit{Euclid}; \citealt{euclid}) will immensely help not only in identifying new GRSs but as well as probing their host AGN and galaxies properties in much finer details.

\subsection{Mid-IR properties} \label{ir}
The optical and UV radiation coming from the accretion disc of the AGN is absorbed by the dusty torus around it and re-emitted in the form of infrared radiation. The WISE mid-IR colour of RLAGNs is a useful parameter to differentiate their host AGN according to their accretion modes, especially in absence of availability of spectroscopic data. Based on this information, \citet{gurkan14} have distinguished the hosts of radio galaxies into sub-sets of HERGs, LERGs, star-forming galaxies (SFGs), narrow-line radio galaxies (NLRGs) and ultra-luminous infrared radio galaxies (ULIRGs) on the WISE colour-colour plot. \citet{Sullivan15}, \citet{mingo16}, and \citet{Whittam18} have confirmed the distinction using other radio samples. Extending the criterion used by \citet{mingo16} for RLAGN to GRSs, \citet{D17,sagan1} have classified the host AGNs of GRGs into high-excitation giant radio galaxies (HEGRGs), low-excitation giant radio galaxies (LEGRGs), star-forming galaxies (SFGs), and ultra-luminous infrared giant radio galaxies (ULIGRGs). The HEGRGs (see figure 3 from \citealt{sagan1}) stand out from the rest with the criterion of W1[3.4 $\muup \rm m$] $-$ W2[4.6 $\muup \rm m$] $>$ 0.5 and W2[4.6 $\muup \rm m$] $-$ W3[12 $\muup \rm m$] $<$ 5.1. The LEGRGs have WISE colours within W1[3.4 $\muup \rm m$] $-$ W2[4.6 $\muup \rm m$] $<$ 0.5 and 0 $<$ W2[4.6 $\muup \rm m$] $-$ W3[12 $\muup \rm m$] $<$ 1.6. The condition of W1[3.4 $\muup \rm m$] $-$ W2[4.6 $\muup \rm m$] $<$ 0.5 and 1.6 $<$ W2[4.6 $\muup \rm m$] $-$ W3[12 $\muup \rm m$] $<$ 3.4 segregates the overlapping region of LEGRGs and star-forming ones from the remaining population. And the ULIGRGs occupy the region with W1[3.4 $\muup \rm m$] $-$ W2[4.6 $\muup \rm m$] $<$ 0.5 and W2[4.6 $\muup \rm m$] $-$ W3[12 $\muup \rm m$] $\geq$ 3.4. Therefore, it is evident that GRGs are not different from SRGs in terms of their excitation states. In the case of the SRQs as well as GRQs, they all assemble in the region occupied by HERGs and HEGRGs in the WISE plot. There is significant overlap between the two and thus, there is no specific criterion to distinguish them based on their WISE magnitudes.

\begin{table*}[ht!]
\centering
\setlength{\tabcolsep}{4.7pt}
\captionsetup{width=5.4in}
\caption{H{\sc{i}} observations of GRSs. References: 1. \citet{Dressel83}; 2. \citet{Morganti09}; 3. \citet{emonts10}; 4. \citet{Chandola13}; 5. \citet{VANGorkom89}; 6. \citet{Conway00}; 7. \citet{Saikia07}. \\
Notes: a: $6.8\times10^7 \rm M_\odot$ of H{\sc i} in emission; b: $<0.41\times10^7 \rm M_\odot$ of H{\sc i} in emission; c: $<2.8\times10^7 \rm M_\odot$ of H{\sc i} in emission. All estimates are from \citet{emonts10}. }
\begin{tabular}{llclcclc}
\hline
 Name & Alt. & Type & redshift & FR class & Col. density & Refs. & Notes \\
     & name  &          & & & $\times 10^{20}$ atom cm$^{-2}$ &  & \\
\hline
J0057+3021 & NGC315 & G & 0.0165 & I-II & 2.5 - 4.5 & 1,2,3 & a \\
J0107+3224 & 3C31  & G & 0.0170 & I & $<$0.6    & 3 & b \\
J0313+4120 & S4     & G & 0.1340 & II & $<$1.68   & 4 &  \\
J0748+5548 & DA240  & G & 0.0357 & II & $<$2.78   & 4 &   \\
J1006+3454 & 3C236 & G & 0.0994 & I-II & 4 - 200   & 5,6 \\
J1111+2657 & NGC3563& G & 0.0331 & I & $<$3.7    & 3 & c  \\
J1147+3501 &        & G & 0.0631 & II & $<$1.96   & 4 & \\
J1247+6723 & VII Zw 485 & G & 0.1073 & II & 0.89 - 2.02 & 7 & \\
J1632+8232 & NGC6251 & G & 0.0247 & I-II & $<$0.95  & 5 \\
\hline
\end{tabular}

\label{tab:HI_obs}
\end{table*}


\subsection{H{\sc i}}
One of the proposed explanations for the giant sizes of the GRSs, is the prolonged availability of fuel for the AGN, allowing it to accrete efficiently and power the radio jets for long periods of time.
Possible evidence of this gas may be probed via observations of neutral atomic hydrogen or H{\sc i} and cold molecular gas. Detecting this gas in emission at moderate or high redshifts is challenging although there has been success in stacking H{\sc i} observations of a large number of galaxies. H{\sc i} in emission has been detected in a number of nearby early-type galaxies \citep[e.g.][]{emonts10}, but most of the attempts to detect H{\sc i} in RLAGN have been via absorption line observations, with the compact radio sources exhibiting the highest detection rates. These studies have been extensively reviewed by \citet{Morganti18}, and a few more recent studies have been summarised by \citet{Odeasaikia21}.

The number of H{\sc i} observations of GRSs are rather few in number (see Tab.\ \ref{tab:HI_obs}). These have so far been largely confined to sources with strong central components. The column densities have been estimated using a spin temperature of 100K and a filling factor of unity and have been taken from the references cited in Tab.\ \ref{tab:HI_obs}. Three of the 9 GRSs listed in Tab.\ \ref{tab:HI_obs} are detected in neutral atomic hydrogen. This is a rather high fraction considering that very few large radio galaxies are detected in H{\sc i} with the column density inversely correlated with the projected linear size \citep{philstrom03,Gupta06size}. 

Two of the early detections of H{\sc i} in absorption have been towards the GRGs NGC315 \citep{Dressel83} and 3C236 \citep{VANGorkom89}. NGC315 has been studied in detail by \citet{Morganti09} and \citet{emonts10}. \citet{Morganti09} suggest that the narrow component which is infalling with a velocity of \simi490 km~s$^{-1}$ is not close to the AGN but due to clouds falling into NGC315 which appears to be in a gas-rich environment. The relatively broad absorption component (FWZI \simi150 km s$^{-1}$) is redshifted by \simi80 km~s$^{-1}$, and could be fuelling the AGN. H{\sc i} has also been detected in emission towards NGC315 and its mass is $\sim6.8\times10^7 \rm M_\odot$ \citep{emonts10}. 3C236, one of the largest GRGs and also a DDRG, with the steep-spectrum core component resolved into an inner double (Fig.~\ref{fig:3c326ims}), has been studied in detail by a number of authors \citep[e.g.][]{Conway00,Schilizzi01,odea01}. VLBI observations of the inner $\sim$\,2 kpc double show that a narrow component (FWHM $\sim$\,50 km~s$^{-1}$) of high opacity occurs at the tip of the eastern component, while much broader (FWHM $\sim$\,200 km~s$^{-1}$) lower opacity gas covers most of the component. They interpret their results to be due to the interaction of the jet with the ISM of the host galaxy. The third source J1247+6723 is again a DDRG where the inner double which has a peaked spectrum is only about 14 pc in size, compared to the overall size of $\sim$\,1200 kpc. The absorption profile obtained with the GMRT shows four components on either side of the systemic velocity \citep{Saikia07}. VLBI spectroscopic observations would be useful to determine the relationship between the different spectral components and the inner double.

Although H{\sc i} has been detected in 3 of the 9 GRSs,
it would be premature to conclude that their large sizes are due to the availability of fuel. Two of these three sources are DDRGs where there appears to be a higher detection of H{\sc i} \citep{Saikia07,chandola10}, and one occurs in an  H{\sc i} rich environment. Clearly, we need 
observations of a much larger sample with high sensitivity to explore these aspects which would be possible with SKA.

\subsection{CO studies}\label{sec:costudies}
The feedback from AGN can affect the gas assembly and the star-formation rate of the host galaxy and in turn properties of the AGN \citep{Croton06}. The transfer of energy from the AGN can also lead to production of outflows and consequently depriving the galaxies of fresh gas and hence quenching star formation \citep{Cicone2014}. 
In a search for cold molecular gas as possible fuel, millimetre-wave observations of a small sample of 3 GRGs were carried by \citet{Saripalli2007}. They provided upper limits using the Swedish-ESO Millimetre Telescope (SEST). In addition to this result, \citet{Saripalli2007} used non-detection results of two other sources from literature (NGC315: \citealt{Braine97} and NGC6251: \citealt{Elfhag96}), and concluded that GRGs lack large ($\gtrsim$10$^{9}$ $\rm M_{\odot}$) molecular gas content in their host galaxies. As part of project SAGAN, \cite{sagan2} carried out a survey of 12 GRGs using the IRAM-30m telescope and reported detection in 3 GRGs and upper limits for the other 9 sources. 
 The detection was primarily in GRGs with host galaxies with discs or spiral morphology. By combining their results with those from previous studies, they find that most GRGs are in the main sequence of galaxies (Figure 7 of \citealt{sagan2}) and their star-formation efficiency is not exceptional. A few studies with deeper observations have detected and discussed in detail the fueling and feedback processes in these sources (e.g. 3C31: \citealt{Lim00} ; 3C236: \citealt{Labiano2013} ; 3C326N: \citealt{Nesvadba2010}; J2345-0449: \citealt{sagan2,Nesvadba21}).

\subsection{AGN properties}
In order to examine the possibility (Sec.\ \ref{sec:models}) of GRSs being powered by a powerful AGN, it is essential to determine its AGN properties like the black hole mass, Eddington ratio and black hole spin. These properties need to be compared with a sample of SRSs too in order to find similarities and dissimilarities. The emergence of large sky area optical and infrared surveys like SDSS and WISE, have not only aided in finding hosts and redshifts of GRSs but have also provided a wealth of spectroscopic and photometric multi-wavelength data. Such data can be used to infer their AGN properties.

\subsubsection{Black hole mass (M$_{BH}$)}
The black hole mass of the host galaxies of the GRGs were estimated in \citep{sagan1} using M$_{\rm BH}$-$\sigma$ relation which translates the effective galactic stellar velocity dispersion ($\sigma$) into the mass of the central black hole. The relation is given as follows:
\begin{equation}
\rm \log \left(\frac{M_{BH}}{M_{\odot}}\right) = \alpha + \beta \log  \left(\frac{\sigma}{200\ {\rm km~s}^{-1}}\right)
\end{equation}
where $\rm \alpha = -0.510 \pm 0.049$ and $\rm \beta = 4.377 \pm 0.290$ \citep{Kormendy_2013}. About 164 GRGs from the GRG-catalogue of \citep{sagan1} have reliable values of $\sigma$ from the SDSS, and hence, the GRG sample was restricted to 164 for studying the black hole mass distribution. To compare the M$_{\rm BH}$ of GRGs with the SRGs, an SRG sample was created from the \citet{bh12rgs} SRG sample after imposing relevant filters (for details see \citealt{sagan1}). 

The two samples were compared in the matched redshift range of 0.034 to 0.534. They observed that both the samples have similar distributions peaking at similar values. Both the distributions have median M$_{\rm BH}$ value of 0.84 $\times$ 10$^9$ M$_{\odot}$. Also, the p-values of 0.86 and 0.46 from the statistical tests of K-S test and WMW test, respectively, establish the same result.

In order to understand which accretion mode is more effective in driving the growth of black holes and thereby, possibly giants, \citet{sagan1} have 
divided the GRG sample from the GRG-catalogue into two accretion states namely LEGRGs and HEGRGs (see Sec.\ \ref{ir}) and compared their M$_{\rm BH}$ properties. They first carried out the comparison of the two classes in the unmatched redshift bin of z $<$ 1 with 94 sources. Further, for the redshift matched bin (0.18 $\leq$ \z $<$ 0.43), the sample consists of 50 sources. In both cases, the results were found to be the same with the LEGRGs having greater M$_{\rm BH}$ values as compared to the HEGRGs as also supported by the statistical tests.

For their sample of \simi32 GRQs, \citet{Kuzmicz12} have estimated the M$_{\rm BH}$ using $\rm H\beta$, MgII, and CIV emission lines, and found the 
values to be in the range of 1.6 $\times$ 10$^9$ M$_{\odot}$ to 29.2 $\times$ 10$^9$ M$_{\odot}$. They found a weak correlation between \mbh (from $\rm H\beta$ and CIV lines) and linear sizes of GRQs and hinted towards a possibility of central engines of GRQs playing an important role for the growth of giants. In their subsequent work  with a larger sample of GRQs (272) and SRQs (367), \citet{kuzmic21} have concluded that GRQs and SRQs are evolved AGNs with no significant difference in M$_{\rm BH}$. \citet{sagan3} also found no difference in M$_{\rm BH}$ for their redshift-matched samples of GRQs and SRQs. To obtain the M$_{\rm BH}$ of GRQs and a comparative sample of SRQs, 
\citet{sagan3} used the SDSS DR14 quasar catalogue of \citet{Rakshit2020}. For the redshift ranges of \z $<$ 0.8, 0.8 $\leq$ \z $<$ 1.9 and \z $\geq$ 1.9, \citet{Rakshit2020} had used the spectroscopic emission lines of $\rm H\beta$, MgII, and CIV, respectively to estimate the black hole masses. \citet{sagan3} applied further filters in addition to the ones suggested by \citet{Rakshit2020} to refine the data of M$_{\rm BH}$. Therefore, the samples were restricted to 191 GRQs and 365 SRQs while comparing in redshift matched bins. The M$_{\rm BH}$ distributions of GRQs and SRQs were found to be similar for redshift matched bins (\z $\leq$ 1.0 and 1.0 $<$ \z $\leq$ 2.45).

When the black hole masses of the four samples i.e., GRGs, SRGs, GRQs and SRQs were compared by \citet{sagan3} in the redshift matched bin of \z $<$ 1.0, it was interestingly observed that the quasar population exceeds the galaxy population with a significant overlap between the two.

\citet{sagan1} using their estimates of M$_{\rm BH}$ and \qj for GRGs, estimated black hole spin (\(\rm \textit{a} \propto \frac{\rm \sqrt{\rm Q_{Jet}}} {\rm B \times M_{BH}} \)) by assuming maximum possible magnetic field i.e., B$_{\rm Edd}$. For their sample of GRGs, they found the median value of 0.055 for the range 0.007$<$ a  $<$ 0.518, where nearly 99\% of their sample have values below 0.3. Further discussion of the result and probable scenarios for giants is given in \citet{sagan1}. They also note that to determine spin of the black hole, the most robust technique available to date is the x-ray reflection, which requires strong sources and deep x-ray observations \citep{reynoldspinnat19}.

\subsubsection{Eddington ratio ($\lambdaup_{\rm Edd}$)}
The Eddington ratio ($\lambdaup_{\rm Edd}$), an indirect estimate of the accretion rate of the black hole, is defined as the ratio of bolometric luminosity (L$_{\rm bol}$) to the Eddington luminosity (L$_{\rm Edd}$) of the black hole and is given as follows:
\begin{equation}
 \rm \lambdaup \equiv \frac{L_{\rm bol}}{L_{\rm Edd}}  
\end{equation}
Here, L$_{\rm bol}$ for the 99 GRGs \citep{sagan1} and the comparative sample of 12998 SRGs from \citep{bh12rgs} was estimated from the following relation where L$_{\rm [OIII]}$ represents the [O{\sc{iii}}]  emission line luminosity:
L$_{\rm bol}$ = 3500 $\times$ L$_{\rm [OIII]}$ \citep{Heckman2004}. And the Eddington luminosity for the two samples was derived from the relation given below:
\begin{equation}
\label{eqn:Ledd}
  \rm   L_{\rm Edd} = 1.3 \times 10^{38} \times \left(\rm \frac{M_{BH}}{\rm M_{\odot}}\right) erg~s^{-1}
\end{equation} 
It is the maximum luminosity of a source when the inward gravitational force of attraction is balanced by the outward force of radiation. The $\lambdaup_{\rm Edd}$ distributions of GRGs and SRGs are found to be different with the SRGs having higher values of $\lambdaup_{\rm Edd}$ as compared to the GRGs.

While comparing the $\lambdaup_{\rm Edd}$ distributions of the two classes of GRGs based on the accretion modes i.e., LEGRGs and HEGRGs, \citet{sagan1} have noticed that the high excitation GRG population is accreting faster than the low excitation GRG class. The two distributions are significantly different with negligible overlap. The conclusion was based on a small sample of 55 sources consisting of 48 LEGRGs and 7 HEGRGs depending on the availability of [O{\sc{iii}}] 
line flux data. However, the results are invariant in redshift matched (0.18 $\leq$ \z $<$ 0.43) and unmatched (\z $<$ 1.0) bins. It has been earlier shown in the literature \citep{hardcastlenat} that LERGs are more dominant in population than HERGs. A similar trend is seen in the case of GRGs in \citet{sagan1}. The findings, overall, indicate the rarer class of HEGRGs are radiatively efficient with a high accretion rate whereas, the LEGRGs lie in the state of radiatively inefficient mode.

\citet{Kuzmicz12} have used optical monochromatic continuum luminosity of 5100 \AA, 3000 \AA, and 1350 $\AA$ to obtain L$_{\rm bol}$ values which were used to estimate $\lambdaup_{\rm Edd}$. They found mean $\lambdaup_{\rm Edd}$ values of GRQs and SRQs to be 0.07 and 0.09, respectively, and concluded that GRQs are evolved or aged sources with declined accretion process. However, \citet{kuzmic21} reported no significant difference in accretion rate of GRQs as compared to SRQs.
\citet{sagan3} carried out a comparative study of $\lambdaup_{\rm Edd}$ values for redshift-matched samples of  GRQs and SRQs by considering the data of L$_{\rm bol}$ from the quasar catalogue of \citet{Rakshit2020}. To estimate L$_{\rm bol}$, \citet{Rakshit2020} used monochromatic  luminosity of the optical lines mentioned above, for the respective $z$ intervals of $z$ $<$ 0.8, 0.8 $\leq$ $z$ $<$ 1.9, and $z$ $\geq$ 1.9. \citet{sagan3} have estimated the L$_{\rm Edd}$ values using Eq.\ \ref{eqn:Ledd}. They reported significant overlap between $\lambdaup_{\rm Edd}$ distributions of both GRQ and SRQ samples but the SRQs have higher values of $\lambdaup_{\rm Edd}$ as compared to GRQs. This accounts for the fact that in spite of having similar M$_{\rm BH}$ distributions, the SRQs are accreting at a higher rate than the GRQs \citep{sagan3}.

\citet{sagan3} have also shown that there is negligible overlap between the distributions $\lambdaup_{\rm Edd}$ when the GRG and SRG samples were compared with the GRQ and SRQ samples, and the latter population is accreting faster than the former one. This implies that the quasar AGNs are highly active and more powerful than non-quasar AGNs \citep{sagan3}.

\section{High energy studies}
X-ray observations of the AGNs allow us to probe their accretion state and obscuration level. 
Using a sample of 14 GRGs selected from soft-$\gamma$ or hard x-ray band, \citet{Ursini18xgrg} presented its broad-band X- ray properties. They showed these GRGs to have higher nuclear luminosities and relatively high $\rm Q_{Jet}$ owing to their radiatively efficient AGN. The $\rm L_{bol}$ estimates of \citet{Ursini18xgrg} are quite high compared to that of \citet{sagan1}, indicating hard x-ray selected GRGs of \citet{Ursini18xgrg} are the extreme examples or tail end of the distribution of GRGs. 

The interaction of relativistic electrons of the radio sources with CMB via inverse-Compton (IC) process boosts the CMB photon energy. Diffuse extended x-ray emission in radio sources arising from the IC-scattered CMB (ICCMB) photons has been detected in several sources (e.g. \citealt{Fabian03,Erlund06}).
The CMB energy density scales as u$_{\rm CMB} \propto ({1+z})^4$, which can exceed the magnetic energy density u$_{\rm B}$ at high $z$ in radio sources. In this case, the cooling would be preferentially via the IC mode rather than the synchrotron cooling as seen in the lobes of GRSs even at intermediate (z$<$1) redshifts \citep{Ishwar-saikia00,Konar2004}.
This can lead to the possibility of detecting only the brighter parts, for example, the hotspots, and missing weaker emission such as the bridge emission. In such cases, it is easy to misidentify the hotspots as independent radio sources.
If the jet activity in such sources stops then they will not only become radio dark, appearing as an inverse-Compton ghost for a while, before also becoming x-ray dark (cf. \citealt{Mocz11}).
Hence, it is likely that this population of `giant radio ghosts' at higher $z$ have been missed till now by current radio telescopes but may be detected with the SKA.
Observationally, \cite{Konar2004} using a sample of a few GRSs ($z<1$) suggested that the prominence of bridge emission decreases with redshift (see Sec.\ \ref{sec:morph}).  There have also been a few detection of ICCMB for high-$z$ GRS, which are intrinsically quite powerful (e.g MRC~2216-206; \citealt{laskar10}, 4C+39.24; \citealt{Erlund08}). \citet{Tamhane2015}, using deep observations with GMRT were able to find a rare steep spectrum relic GRG (J021659$-$044920) at $z$\simi1.3 with x-ray emission coincident from the radio lobes (ICCMB). They did not detect the radio core, which indicates that the radio activity may have ceased, which is also supported by the steep spectrum nature of the radio emission from the lobes.

Based on their observational data, \citet{nath10} and \citet{Mocz11} developed a model for double-lobed sources of FRII type in the x-ray sky (using information from the \textit{Chandra} x-ray Deep Field North survey). \citet{nath10} predicted a space density of \simi25 deg$^{-2}$ above an x-ray flux limit of \simi3 $\times$10$^{-16}$~erg~cm$^{-2}$~s$^{-1}$ for such sources. On the other hand, \citet{Mocz11} 
predicted that about 10-30\% of the observable double-lobed structures in the x-ray are IC ghosts. They also predict that such sources have higher space densities at $z\geq2$ compared to the x-ray detected galaxy clusters.

\section{Environment}
Detailed environmental studies of GRSs require deep x-ray and optical spectroscopic observations of the surrounding medium/objects, which is expensive in terms of telescope time, especially for a large sample of sources. Hence, such detailed studies have been restricted to a few sources \citep[e.g.][]{ravi08,chen11ngc6251,chen11da240,Pirya12,chen12ngc315,chen124c73,Malarecki13}.
\citet{ravi08} carried out a large optical spectroscopic survey around the GRG~0503$-$286 \citep{saripalli86}. GRG~0503$-$286 is a powerful FRII type GRG with asymmetry, where its radio core is closer to its northern lobe than the southern lobe. North of the GRG, there are multiple galaxy groups at similar redshifts indicating a denser environment, whose effects are clearly seen in the growth of GRG~0503$-$286 towards the north. Using a sample of 19 GRGs and spectroscopic observations of \simi9000 galaxies, \citet{malarecki15} showed that in asymmetric GRGs, the shorter lobe is on the side of denser environments. \citet{Pirya12} examined the distribution of galaxies in the vicinity of 16 large radio sources almost all of which are GRGs. They found that in the GRG 3C326, which has the highest arm-length ratio in their sample, the shorter arm appears to interact with a group of galaxies which form part of a filamentary structure. They also reported that while most sources appear to occur in regions of low galaxy density, the shorter arm is brighter in most cases. This suggests that there are asymmetries in the intergalactic medium which may not be apparent from the distribution of galaxies \citep{Pirya12}.  Recently, \citet{Dabhade22}, reported the X-shaped nature of GRG~0503$-$286, where they observe an inversion-symmetric bending of the lobes with a radio emission gap of \simi200 kpc between them. Based on its radio morphology, source energetics, and the environment, they attribute the observed peculiar feature of the emission gap to the presence of gaseous layer or WHIM-filled sheet (Warm-hot-intergalactic medium; \citealt{dave01}) in the cosmic web. Deeper x-ray observations of the environments of such sources are needed to substantiate the suggested possibility.

As mentioned in Sec.\ \ref{sec:models}, one of the most favoured explanations for the large sizes of GRSs is their growth into possibly sparse environments. Studies of samples of GRSs using radio data \citep{mack98,Schoenmakers_spec,Stuardi20} have shown that these sources preferentially grow in under-dense environments (thermal electron densities $\rm <10^{-5}~cm^{-3}$). It is essential to examine the environments of GRSs using other methods (e.g. X-rays and optical spectroscopy) and also compare with controlled samples of smaller radio sources.

In contrast to the above, using the large compilation of GRS samples, \cite{PDLOTSS,sagan1} showed that at least 10-20\%  GRGs ($z<0.7$) reside in the denser environment at the centres of galaxy clusters. They also noted that these sources avoid the very massive rich clusters. A similar result was also noted by \cite{Komberg09} earlier.
However, it is still unclear what fraction of GRSs reside inside the galaxy cluster (within virial radius: R$_{200}$ or R$_{500}$)  and if there is any preference in terms of their location from the centres of galaxy clusters.

\citet{lan21} using a sub-sample of GRGs from the LoTSS DR1 sample of \citet{PDLOTSS} along with two control samples, a radio control sample and an optical one, explored possible relationship with galaxy surface density and relationship with large-scale structure. They reported that the locations of GRGs and control samples relative to  large-scale cosmic web filaments are similar. Hence, they too conclude that GRGs do not reside preferentially in low-density environments in the context of large-scale structure, which possibly could have favoured their growth to megaparsec scales. They also find that the properties of satellite galaxies around GRGs from 20~kpc to 2~Mpc are similar to the properties of satellite galaxies in the control samples. Thus the current period of activity in the GRGs appear to have little influence on the properties of the surrounding galaxies.

\section{Unification}
Orientation-based unified schemes where the observed properties of AGN are due to a combination of orientation and relativistic motion of jets have had a fair degree of success in understanding their properties and reducing the apparent diversity of AGN. These have been reviewed extensively in the literature (e.g. \citealt{Antonucci93,unification95,Tadhunter16}). For RLAGN there are two different schemes for the FRII and FRI sources as the extended structure should not depend significantly on orientation. For the FRII radio sources with strong emission lines, the narrow-line radio galaxies are believed to be close to the sky plane, while those with broad lines and quasars are inclined at small angles to the line of sight. For the FRI radio sources, the weak-lined radio galaxies are close to the sky plane, while their counterparts inclined at small angles to the line of sight are the BL Lac objects. Although these schemes have had a reasonable degree of success, it does not account for the entire diversity of properties of RLAGN which may also depend on the accretion mode, properties of the supermassive black holes, host galaxy properties and the environment (cf. \citealt{Tadhunter16,Mingo22}). In this section we focus largely on FRII radio sources, as the vast majority of GRSs have been classified to be FRII sources \citep{sagan1}.

  \begin{figure}
    \includegraphics[scale=0.31]{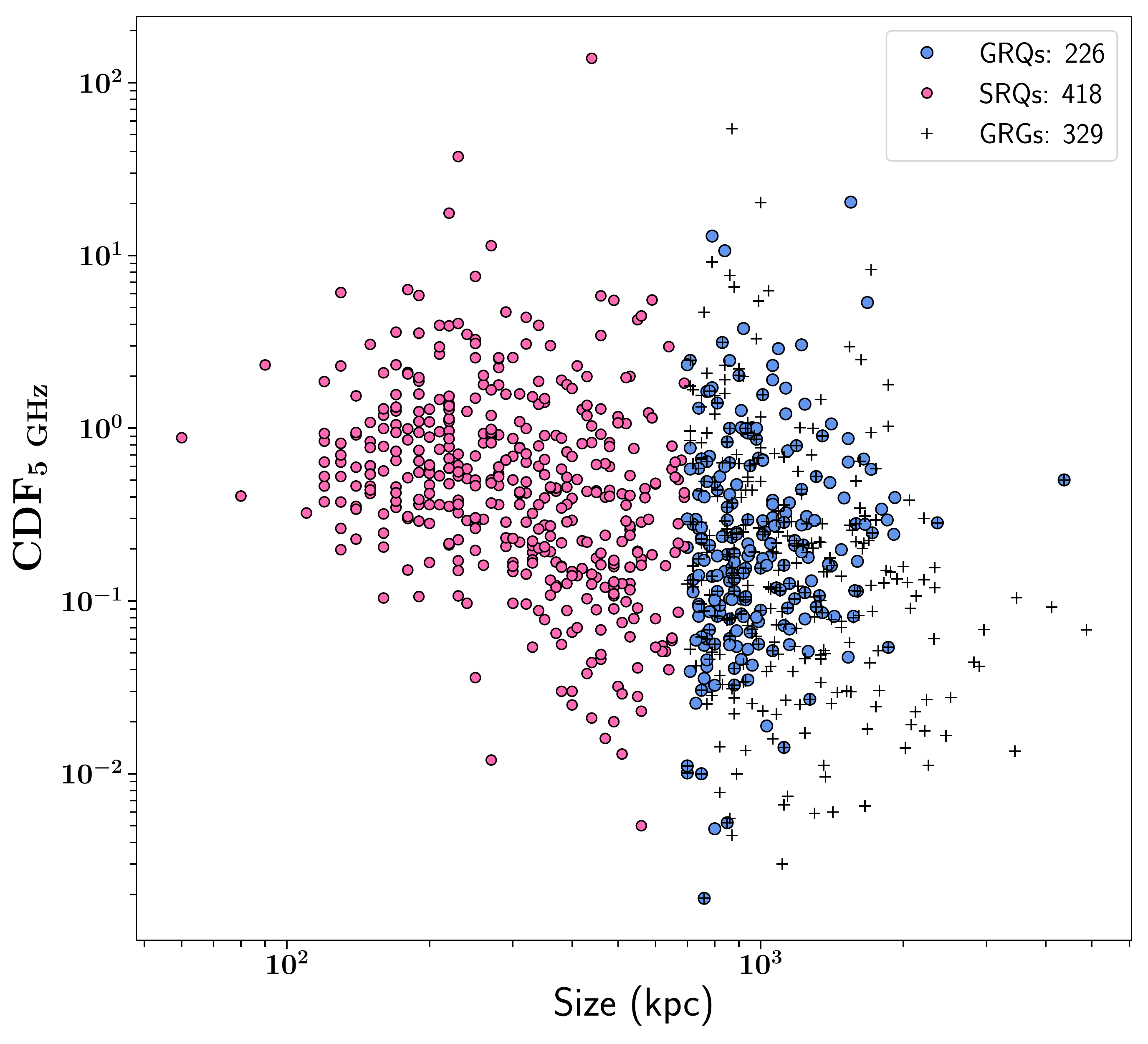}
    \caption{The plot shows CDF as a function of the projected linear size of SRQs, GRQs, and GRGs, where the quasar data is taken from \citet{sagan3} and the GRG data from \citet{sagan1}.}
    \label{fig:cdf_size}
\end{figure}

\subsection{Core dominance factor}
Nevertheless, in RLAGN where the jets are moving at relativistic speeds close to that of light in the nuclear regions, relativistic beaming will play an important role in the prominence of the radio core.  Hence,
the core dominance factor (CDF or f$\rm _c$) is a significant tool to probe the orientation of jet axis with respect to our line of sight. Due to relativistic beaming, the radio emission from the core is enhanced in comparison with the emission from the extended regions. Therefore, there should be an anti-correlation between CDF and projected linear size as suggested by AGN unification scheme.  The CDF is defined as the ratio of core flux density to the flux density observed from the extended regions of sources excluding the core. The rest-frame flux density $\rm S_{\nu_{rest}}$, the observed flux density $\rm S_{\nu_{obs}}$, spectral index $\alpha$ and redshift ($z$) are related as follows:
\begin{equation} 
   \rm  S_{\nu_{rest}} = S_{\nu_{obs}}(1+z)^{(\alpha -1)}\left(\frac{\nu_{rest}}{\nu_{obs}}\right)^{(-\alpha)}
 ,\end{equation}
 where  $\rm \nu_{rest}$ and $\rm S_{\nu_{obs}}$ are the rest-frame and observed flux densities, respectively \citep{sagan3}. 
 
In an early attempt to develop the unified scheme, \citet{kapahisaikia82} reported such a trend by showing that as the size of the sources increases, the core prominence decreases for a sample 78 well-observed radio quasars. There have been a large number of studies of using the prominence of the core as an indicator of orientation of the jet axes to the line of sight \citep[e.g.][]{Browne87,Ishwar-saikia00,marin16,2020Maithil}. Along similar lines, in a comparative study of properties of SRQs and GRQs, \citet{sagan3} have confirmed that the core prominence of SRQs is almost twice  that of GRQs for significantly larger samples.
In their study of GRQs, \citet{sagan3} have obtained CDF at the rest-frame frequency of 5~GHz. The GRQs in their sample have median CDF values of 0.44 and 0.98 for the lower and upper redshift bins, whereas for the SRQs the respective values are 0.87 and 1.63 respectively. It indicates that GRQs are inclined at larger angles with respect to our line of sight than the SRQs. 

One can also examine whether the GRQs have more prominent cores than the GRGs, as one  might expect in the unified scheme. \citet{Konar2008} observed a sample of 10 GRGs for which the CDF values at the rest-frame frequency of 8 GHz range from 0.004 to 0.14 with a median value of 0.04. We have computed the CDF values for the GRGs from the GRG-catalogue of \citet{sagan1} at the rest-frame frequency of 5~GHz as shown in Fig.~\ref{fig:cdf_size}. For this, we have obtained the 1.4 GHz observed flux densities from the FIRST survey source catalogue \citep{Helfandfirst}. The median value of core prominence for GRGs is found to be 0.15, significantly smaller than that for quasars, consistent with the expectations of the unified scheme.

\subsection{Jets and superluminal radio sources}
In the unified scheme for radio galaxies and quasars, one would expect quasar jets to be more asymmetric and possibly exhibit superluminal motion when inclined at small angles to the line of sight. The fraction of GRSs with detected jets is rather small. More sensitive high-resolution observations should enable us to detect jets in a much larger number of sources. Here we merely highlight a few sources and discuss their consistency with the unified scheme. 

Two of the GRQs with prominent jets are 4C34.47 \citep{Barthel89vlbi} and 4C74.26 \citep{riley89}.
4C34.47 has a prominent large-scale jet without any counter jet with the flux density ratio $>$10. VLBI observations of the core show a nuclear jet exhibiting superluminal motion indicating that this component lies within 44$^\circ$ to the line of sight \citep{Barthel89vlbi}. The nuclear jet in 4C74.26 is also aligned with the large-scale jet. Although there are no clear features to monitor for the motion of the nuclear jet, the flux density ratio suggests an angle of inclination $<$ 49$^\circ$ \citep{pearson92}. One would expect the jets in GRGs to be inclined at larger angles. The jets in the GRG 3C31 have been modelled in detail by \cite{laing02} who suggest the jet to be inclined at $\sim$52$^\circ$ with the jets decelerating as these travel outwards. However, two of the well-studied GRGs whose jets have been studied and modelled in detail yield inclination angles of $\sim$38$^\circ$ for NGC315 \citep{Canvin05} and $\sim$32.5$^\circ$ for NGC6251 \citep{laing15jetq}.
These appear to be smaller than GRQs contrary to the expectations of the unified scheme. However,
these two GRGs have very prominent cores and may not be representative of all GRGs. More sensitive observations as well as detailed modelling of large samples of sources which will be possible from observations with the SKA will help our further understanding of these aspects.

\section{GRSs as probes and their association with other processes in the Universe}
\subsection{GRSs as probes}
Using multi-frequency observation from 1 to 10 GHz along with polarisation measurements for 3C~445 and PKS~0634-20, \citet{Kronberg86} demonstrated the use of GRGs as sensitive probes of arcminute scale variations of foreground Galactic RM. Their work also estimated the IGM density and temperature. This is the earliest work where GRGs were used as probes for other processes. More recently, this was demonstrated by \citet{Stuardi20} with higher precision owing to the use of low-frequency measurements from LOFAR. This will be further explored in the SKA-era astronomy \citep{Vacca15}.

The x-ray observations have revealed the presence of a halo of hot gas around galaxies, especially the ones in the centres of galaxy clusters, which causes depolarisation in radio-emitting components closer to the parent galaxy in RSs \citep{Strom88,Johnson95,ishwar98}. In other words, the effect of Faraday depolarisation is much higher towards the centre (or parent galaxy) of the radio source, as it is known that magnetic field strength and ionised gas density decrease with radius in galaxy groups and clusters \citep{laing08,Bonafede10}. The correlation of depolarization asymmetry with jet-sidedness, usually referred to as the Laing-Garrington effect, is usually attributed to a halo of hot gas associated with the radio source \citep{Garrington1991}. \citet{GKNATH97} proposed the presence of extended magneto-ionic disks in elliptical galaxies hosting quasar type AGN to explain the asymmetry. The GRGs, which are possibly growing in under-dense environments, with their lobes well beyond the gaseous extents of their parent galaxies are expected to be highly polarised even at low frequencies (see also Sec.\ \ref{sec:pol}). \citet{Stuardi20} convincingly demonstrated the same using \citet{PDLOTSS} LoTSS GRS sample. Hence, the high degree of polarisation in GRGs can make them potential polarisation calibrators for weaker sources.

GRGs with their polarised lobes expanding in the IGM or filaments or cosmic voids
can also be used as probes of the intergalactic magnetic field. \citet{Sullivan19}, using GRG-J1235+5317 suggested that the gas density-weighted intergalactic magnetic field (IGMF) strength is \simi0.3 $\muup$G. Similarly, \citet{Stuardi20} using a sample of GRSs clearly demonstrated its potential use for studying magneto-ionic properties of large-scale structures.

GRSs can also be used as probe of the low density (\simi10$^{-5}$ to 10$^{-6}~ \rm cm^{-3}$) WHIM to study the `missing baryons' problem. A few studies of GRSs have shown the large extent of the sources possibly allow them to interact with WHIM \citep{mack98,chen11ngc6251,Malarecki13,malarecki15}. While detection of WHIM is observationally quite challenging, GRGs can possibly act as indirect probes \citep{ravi08}.

\subsection{Possible association with cluster radio relics}
Apart from GRSs, the other radio emission extending to Mpc scale is from the radio relics associated with the galaxy cluster merger shocks \citep{Rottgering97_3667,Bagchi06_3376,vanWeeren19}. The seed electron population (re-acceleration) needed for generating massive shock relics is still being investigated. GRSs, owing to their large sizes are good candidates as their lobes cover hundreds of kpc volume.
There have been a few examples (Abell 3411-3412; \citealt{Weeren17}, A3376; \citealt{Chibueze22}), where a connection between a RG and the radio relic has been observed, providing evidence on how radio galaxies can supply fossil electrons needed for creation of radio relics.

\subsection{Sunyaev-Zel'dovich effect}
When the CMB photons are boosted to higher energies via inverse Compton scattering with the non-thermal population of electrons, it leads to the shifting of the CMB spectrum and this effect is known as the \textit{Sunyaev-Zel'dovich} effect \citep{sz70,sz80}.
A few studies have attempted to study this effect using RSs and GRSs (e.g. \citealt{Yamada10,Colafrancesco11,Colafrancesco13}). The role of GRSs in contributing towards anisotropies in the CMB  and hence, acting as possibly a contaminant towards \textit{Sunyaev-Zel'dovich} surveys has been suggested by \citet{blundell06} and \citet{Erlund06}, which is based on the work of \cite{esnsslin00} for radio emission in general from extragalactic sources. GRSs, owing to their enormous size like galaxy clusters could make a significant contribution.

\subsection{Sites for acceleration of ultra high energy cosmic rays?}
It is likely that GRGs with their favourable large lobe sizes of hundreds of kpc with ongoing turbulent accelerations in hotspots are possible sites for the acceleration of charged particles to very high energies. These are commonly referred to as the ultra high-energy cosmic rays (UHECRs; \simi10$^{18}$ - 10$^{21}$ eV) or particles \citep{Hillas84}. A few studies with semi-analytical models \citep{Blandford00,Fraschetti08,Matthews18} have suggested jetted AGNs to be the likely origin for UHECRs, where shocks within the energetic jets and lobes can accelerate the particles to such extreme energies. Particularly, figure 1 from \citet{Hardcastle10} shows that large-sized RGs or GRGs are more prone to be likely the sites for the acceleration of UHECRs. The association of GRGs with efficient particle acceleration was first discussed by \citet{kronberg04} in detail using the theory of collisionless magnetic reconnection and later by \citet{hardcastle09} and \citet{Sullivan09} for Centaurus A. In the context of high-energy particles it is interesting to note that the IceCube Neutrino Observatory\footnote{\url{https://icecube.wisc.edu/about-us/overview/}} has reported neutrino emission from the direction of the blazar TXS~0506+056 \citep{IceCubesci18}.


\section{Concluding remarks and scope with SKA} \label{sec:concl}

The number of known GRSs, although constituting only a small fraction of RLAGN, have increased dramatically to  \simi3200 with the recent surveys with LOFAR and ASKAP \citep{PDLOTSS,Andernach21,oeidr2}. High sensitivity and suitable resolution especially at low radio frequencies have contributed to this very significant increase. These numbers will further increase dramatically
when the SKA becomes available \citep{peng15}. In this section, we briefly summarise our current understanding and possibilities of
future work. 

Let us first consider the models which have been suggested to understand the formation of GRSs. Both dynamical and spectral age studies of GRSs and SRSs demonstrate that GRSs are older with ages approaching several times 10$^8$ yr
(Sec.\ \ref{sec:specage}, Tab.\ \ref{tab:specage}, and Fig.\ \ref{fig:dagesize}). The distribution of GRSs in the P-D diagram also shows that these are the oldest systems with theoretical models suggesting that a wide range of jet powers starting from $\sim$10$^{35}$ W can lead to the formation of GRSs (Sec.\ \ref{sec:models} and Fig.\ \ref{fig:pd}). 

Among the different scenarios suggested for the formation of GRSs, one possibility has been that these occur in low-density environments. However, a number of studies
have shown \citep[e.g.][]{Komberg09,PDLOTSS,sagan1,Tang20} that at least 10-20\% of GRSs reside in centres of galaxy clusters as BCGs. Hence, a sparse environment exclusively cannot be the only reason for their enormous sizes. Most GRSs do tend to occur in regions of low galaxy density or low density environment \citep[e.g.][]{Stuardi20}. However, even among these sources the brighter component often tends to be closer \citep[e.g.][]{Pirya12}, suggesting that there are asymmetries in the gas distribution in the vicinity of the GRSs which may not be reflected in the distribution of galaxies. Sensitive x-ray observations could help reveal the gas distribution of these sources.

Another possibility is that the AGNs of GRSs are extremely powerful, capable of producing jets which propagate such large distances \citep[e.g.][]{gk89}. This also does not stand up to observational scrutiny. GRSs are seen over a wide range of jet powers \citep[e.g.][]{sagan1,gurkan22}. Also, \citet{Ishwara1999} found that the core strength for a sample of GRGs is similar to that of smaller-sized sources matched in extended radio luminosity. This suggests that GRGs do not necessarily have a more powerful AGN. Recurrent nuclear jet activity has also been suggested to help aid these sources to grow to Mpc scales \citep[e.g.][]{ravi96}. However, only a small fraction of GRSs are known to exhibit evidence of recurrent activity as discussed in Sec.\ \ref{sec:rejuv} 

The formation of these largest structures by a single galaxy is still poorly understood, and it is possible that a combination of different factors may be responsible for each GRS. As an example, let us consider the two largest GRGs, \textit{Alcyoneus} with a size of 4.98 Mpc \citep{oeip122} and J1420$-$0545 with a size of 4.87 Mpc \citep{Machalski_2008}. \textit{Alcyoneus} is identified with a low-excitation elliptical galaxy with a stellar mass of  $2.4\pm0.4 \times 10^{11} \rm M_\odot$ and a supermassive black hole with a mass of $4\pm2 \times 10^{8} \rm M_\odot$. Both these values are at the lower end of the distributions for GRGs. Thus neither a very massive galaxy nor black hole nor high radio power seems necessary to produce the largest GRG \citep{oeip122}. The source may be evolving in a region of low density. The GRG J1420$-$0545 associated with a typical elliptical galaxy has been modelled to evolve in a low-density environment with a high expansion speed \citep{Machalski_2008}. SKA should help discover more such giant sources whose detailed studies would help further clarify the formation of such structures.

Recent deep observations of the sky with LOFAR \citep{PDLOTSS,oeip122,oeidr2,Simonte2022} and MeerKAT \citep{Delhaize21} have clearly demonstrated that not only fainter GRSs can be found but also new very low surface brightness features (e.g. collimated synchrotron threads) associated with radio sources or GRSs \citep{Ramatsoku20,cotton20} can possibly be observed. Hence, it has opened a new parameter space to be explored and studied. The data from the SKA is most likely to detect extremely low surface brightness sources in less amount of time. Also, it may lead to the possible finding of many high redshifts (z $>$ 3) GRSs. It will allow us to study and understand the growth of GRSs in high-z environments where the CMB density is high and their possible interplay via the IC process. Such studies will also enable us to use GRSs as a probe of the Warm-hot-intergalactic medium (WHIM) to study the `missing baryons' problem.

Besides clarifying the frequency of occurrence of GRSs in the entire population of RLAGN, a large population of low power GRSs would provide further insights into understanding these sources. The SKA will enable us to study both nuclear jets via VLBI observations \citep{agudo15} and large-scale jets in both total intensity and linear polarization to address many of the outstanding questions on jet physics \citep{laing15jetq}. The fraction of GRSs with well defined radio jets as defined by \citet{Bridle84} is rather small. The SKA will enable detection of radio cores and jets in most GRSs which will enable us to test the unification schemes for this class of AGN. Besides continuum observations, detection of H{\sc i} in both emission and absorption in a large number of GRSs including those with recurrent activity will help explore the dynamics of the gaseous component in the host galaxy and availability of fuel for the AGN. The high sensitivity will perhaps reveal many more sources with recurrent activity even in the weaker sources, and help us understand its impact on the host galaxies and galaxy evolution. The advent of the WSRT, VLA and more recently GMRT, LOFAR, MeerKAT, JVLA, and ASKAP have had a significant impact in our study of GRSs and there has been a resurgence of interest in the field. The SKA will be the next major leap forward.

\section*{Acknowledgements}
We thank the anonymous referee for her/his valuable comments and suggestions. We are thankful to Naoki Isobe (JAXA) for providing us with the data from his papers. We thank Martijn Oei for sharing his results from LoTSS DR2.
We thank Marek Jamrozy, Agnieszka Kuźmicz and Sagar Sethi for their valuable feedback. We also thank Shishir Sankhyayan for his help. We gratefully acknowledge the use of Edward (Ned) Wright's online Cosmology Calculator. This research has made use of the VizieR catalogue access tool, CDS, Strasbourg, France. The original description of the VizieR service was published in \citet{vizier}. This research has made use of the NASA/IPAC Extragalactic Database (NED), which is funded by the National Aeronautics and Space Administration and operated by the California Institute of Technology. \textit{Hubble space telescope} images have been obtained from the \textit{Hubble} legacy archive :\url{http://hla.stsci.edu}. We acknowledge the use of the data from the legacy surveys: \url{https://www.legacysurvey.org/acknowledgment/}. We acknowledge that this work has made use of  \textsc{astropy} \citep{astropy}.


\bibliography{GRS_review}

\end{document}